\documentclass[apj]{emulateapj}

\newcommand{\kev}{keV}
\newcommand{\etal}{et al.}
\newcommand{\nh}{$N_{\mathrm{H}}$}
\newcommand{\chandra}{\textit{Chandra}}
\newcommand{\xmm}{\textit{XMM-Newton}}
\newcommand{\spitzer}{\textit{Spitzer}}

\newcommand{\cloudy}{Cloudy}
\newcommand{\fe}{Fe~K$\alpha$}

\slugcomment{Accepted by the \apj\ on Aug.\ 31, 2006}

\shorttitle{An Evolving AGN Unified Model: X-ray to Mid-IR}
\shortauthors{Ballantyne \etal}

\usepackage{times}
\usepackage{graphicx}

\begin{document}

\title{Does the AGN Unified Model Evolve with Redshift?\\ Utilizing the X-ray
  Background to Predict the Mid-Infrared Emission of AGN}



\author{D. R. Ballantyne\altaffilmark{1}, Y. Shi\altaffilmark{2},
  G. H. Rieke\altaffilmark{2}, J. L. Donley\altaffilmark{2},
  C. Papovich\altaffilmark{2} and J. R. Rigby\altaffilmark{2}}
\altaffiltext{1}{Department of Physics, University of Arizona, 1118 East 4th
  Street, Tucson, AZ 85721; drb@physics.arizona.edu}
\altaffiltext{2}{Steward Observatory, University of Arizona, 933
  N. Cherry Avenue, Tucson, AZ 85721}

\begin{abstract}
Deep X-ray surveys by \chandra\ and \xmm\ have resolved about 80\% of
the 2--10~\kev\ cosmic extragalactic X-ray background (CXRB) into
point sources, the majority of which are obscured AGN. The obscuration
might be connected to processes within the host galaxy, possibly the
star-formation rate. Here, we use the results of CXRB synthesis
calculations as input to detailed Cloudy simulations in order to
predict the evolution of AGN properties at several mid-IR
wavelengths. Computations were performed for three different
evolutions of the AGN type 2/type 1 ratio between $z=0$ and $1$: where
the ratio increased as $(1+z)^{0.9}$, as $(1+z)^{0.3}$ and one with no
redshift evolution. Models were calculated with the inner radius of
the absorbing gas and dust at 1~pc or at 10~pc. Comparing the results
of the calculations to combined X-ray and \spitzer\ data of AGN shows
that the predicted spectral energy distributions are a good
description of average AGNs found in the deep surveys. The existing
data indicates that the mid-IR emission from an average
AGN is best described by models where the attenuating material is
$\sim 10$~pc from the central engine. We present the expected
\spitzer\ cumulative number count distributions and the evolution of
the total AGN (type 1 + type 2) luminosity function (LF) between $z=0$
and $1$ at rest-frame 8\micron\ and 30\micron\ for the three
evolutionary scenarios. The mid-IR AGN LF will be an excellent tool to
measure the evolution of the covering factor of the gas and dust from
$z \sim 0$ to $1$.
\end{abstract}


\section{Introduction}
\label{sect:intro}
The earliest optical investigations of active galactic nuclei (AGN)
identified two major classes of sources: type 1 AGN exhibited both
broad (FWHM $\sim 10^4$~km ~s$^{-1}$) permitted lines and narrow (FWHM
$\lesssim 10^3$~km ~s$^{-1}$) forbidden lines in their spectra, while
type 2 AGN presented only the narrow lines \citep{kw74}.  Unification
models of AGN propose that this difference can be simply explained by
an obscuring medium that surrounds a basic black hole/accretion disk
system, and that different lines-of-sight into and through this
obscuration give rise to the variety of observational properties
witnessed in AGN phenomenology \citep[e.g.,][]{ant93}. The obscuring
medium is often visualized as a geometrically thick torus comprised of
dust and molecular gas and of order a parsec in size, situated between
the broad-line and narrow-line regions
\citep[e.g.,][]{pk92a}. Additional obscuration is likely to occur in
larger structures associated with the host galaxy
\citep[e.g.,][]{mr95}. The net unobscured view from the AGN must
encompass only about 20\% of the sky to explain the roughly 4:1 ratio
of obscured type 2 to unobscured type 1 AGN in the local Universe
\citep[e.g.,][]{mr95,hfs97}. This picture is probably consistent with
the recent finding from Sloan data by \citet{hao05} that the type
2/type 1 ratio is close to unity, because about half of the AGN appear
to be well hidden in their host galaxies and would not be seen by
Sloan but can only be detected with very deep spectra such as those
used by \citet{mr95} and \citet{hfs97}. The unification model explains a number
of observations. For example, the near infrared (IR) emission of many
AGN shows a break near 1\micron\
\citep[e.g.,][]{rie78,neu79,bar87,san89}, signifying dust that is at
or near its sublimation temperature and hence close to the
AGN. Recent \chandra\ and \xmm\ X-ray observations have found that
narrow Fe K$\alpha$ lines are common in both type 1 and type 2
objects, which implies the presence of cold, low-velocity gas
intercepting a significant fraction of the nuclear emission
\citep{yp04,nan06}.

While the existence of the absorbing gas around AGN is well
established, the geometry, distribution, origin, and evolution of the
material is largely unknown. Observations of the soft X-ray absorption
can measure the equivalent neutral column density of the absorbing gas
along the line of sight, \nh. There is a good correspondence (at the
$\sim$ 80\% level; \citealt{toz06}) between the sources with \nh\ $\ge$ 10$^{22}$
cm$^{-2}$ and those optically identified as being of type 2. This
correspondence implies that, on average, the X-ray absorbing gas does
lie outside the broad-line region. However, observations of \nh\
variability on week- to year-long timescales \citep{ris02,ris05} argue
against a simple geometry. Further evidence for a complicated geometry
is found in the mid-IR.  The IR spectra of most AGNs (in $\nu f_\nu$
units) peak between 3 and 10\micron, and roll over at $\sim$
100\micron\ \citep{an03,kur03}. This behavior can be described by
thermal emission by dust ranging in temperature from $\sim 40$~K to
the sublimation temperature of 1000--2000~K. Part of the cooler
emission may be heated by star formation rather than the AGN
\citep[e.g.,][]{mr95,sch06}. Interestingly, the spectral energy
distributions of type 1 and type 2 AGN are similar in the mid-IR
($\lambda > 5$\micron) \citep{sm89,fad98,kur03,ah03,lutz04,rig04}, in
conflict with the predictions of many compact torus models
\citep{pk92a,gd94,err95,gra97}. All of these observations argue for a
more distributed or complex geometry to the absorbing material around
the AGN \citep{kur03}.

If the obscuring material in AGN is distributed in a complex manner,
then large object-to-object variations in properties would be
expected, complicating the interpretation of both the AGN
characteristics and the relationship to the host galaxy (if
any). Progress can be made, however, by determining the ratio of type
2 to type 1 objects (hereafter denoted as $R$) at differing AGN
luminosities and redshifts. Under the unified models, $R$ is related
to the covering factor of the absorbing gas and dust, so changes in
$R$ give clues to the nature of the obscuring structures. For
instance, the simplest unified model, that of an unevolving torus,
predicts no evolution of $R$. However, over the last few years,
surveys at multiple wavelengths have noted that the fraction of type 1
objects increases with AGN luminosity
\citep{ueda03,bar05,simp05,hao05}. This trend may simply be a
selection effect \citep{trei04}, because it is easier to find type 1
high luminosity AGN (i.e., quasars) with their broad optical lines
than type 2 objects, which many models predict to be deeply buried
\citep[e.g.,][]{hopk05}. Moreover, at high redshifts dilution from
galaxy light can overwhelm the narrow and relatively weak lines of a
type 2 AGN, increasing the difficulty of identification
\citep{mfc02}. Alternatively, the luminosity dependence of $R$ may be
real, and imply that some part of the obscuring material is close
enough to the central engine to be affected by high luminosities,
perhaps through dust sublimation \citep{law91} or radiative
acceleration of magnetocentrifugal winds \citep{kk94}.

There are also hints that $R$ increases with redshift at moderate AGN
luminosities. Both \citet{bar05} and \citet{laf05}, in their analyses
of deep X-ray surveys, found indications for an increase in the type 2
fraction from $z = 0$ to $z \sim 1$. The sources discovered in these
observations account for a large fraction of the cosmic X-ray
background (CXRB) in the 2--10~\kev\ range \citep{mush00,bh05,wor05},
whose spectrum can be fitted only if $R \sim 3-4$ \citep{gill04}.

The implication of these observations is that if the obscuring medium
around an AGN evolves with redshift then it must be connected in some
way to the evolution of the host galaxy. The redshift distribution of
the CXRB sources peaks at $z \sim 1$ \citep{toz01,bar02,bar05},
interestingly close to where the cosmic star-formation rate (SFR)
density also reaches its maximum \citep{hopk04}. This coincidence in
redshift distribution suggests a possible connection between the
absorbing gas and dust around an AGN and the SFR within the host
galaxy (already seen in the Seyfert~1 NGC~3227;
\citealt{dav06}). With this motivation, \citet{bem06} determined that
a $R$ that evolves with both luminosity and $z$ (constrained by the
type 1 fractions observed by \citealt{bar05}) can fit the observed
CXRB spectrum and X-ray number counts. However, the CXRB itself
provides little constraint on the evolution as both rapid, slow or
zero \citep[e.g.,][]{tu05} evolution can account for the data. A
measurement of the true extent of any $R$ evolution will provide vital
information for models of galaxy and black hole evolution over this
redshift range. It is therefore important to search for other
observational signatures that can differentiate among various
evolutions of the AGN obscuring material.

The natural next step is to consider the evolution of AGN, in
particular obscured AGN, in the mid-IR where the absorbed radiation
will be re-emitted.  With the recent launch of the \spitzer\ Space
Telescope, deep sensitive surveys of AGN in the mid-IR are now being
performed, and are finding many high-$z$ ($z \gtrsim 0.5$) AGN
\citep{lacy04,rig04,fra05,stern05,ah06,barm06,brand06}. Therefore,
this paper presents predictions for the mid-IR number counts and
luminosity functions (LFs) for three different evolutions of the type
2/type 1 AGN fraction between $z=0$ and $1$. All three evolutions are
chosen to fit the CXRB, and therefore the mid-IR predictions are
consistent with the current best X-ray constraints on AGN properties
and evolution. In contrast to most previous work \citep[e.g.][]{silva04}, the IR predictions
are not based on observed spectral energy distributions (SEDs) but are
computed using the photoionization code \cloudy. This method has the
advantages of being able to directly relate the X-ray properties to
other wavelengths and to allow interesting physical properties of the
obscuring medium to be varied and possibly constrained.

We begin in the next section by reviewing the CXRB results of
\citet{bem06}, in particular the three evolutions of $R$ that will be
investigated here. The following section (\S~\ref{sect:calcs})
describes the \cloudy\ models in detail, including a description of
our assumptions, and how we average individual spectra together before
estimating observables. Sect.~\ref{sect:tests} then checks how well
these model SEDs compare against the observed AGN X-ray and mid-IR properties. Sect.~\ref{sect:agn} presents the number counts and LFs for
the three AGN evolutions and compares the predictions to the available
data. In Sect.~\ref{sect:sf} we attempt to take into account the
effects of star formation in the model spectra and present revised
predictions. Finally, we discuss our results in
Sect.~\ref{sect:discuss} and summarize in Sect.~\ref{sect:summ}.

This paper assumes the standard first-year \textit{WMAP}
$\Lambda$-dominated cosmology: $H_0=70$~km~s$^{-1}$~Mpc$^{-1}$,
$\Omega_{\Lambda}=0.7$, and $\Omega_{m}=0.3$ \citep{spe03}.

\section{Review of CXRB Models}
\label{sect:cxrbreview}
When fitting the CXRB spectrum, a key parameter is the ratio of type 2 to type
1 AGN at a given 2--10~\kev\ X-ray luminosity $L_{X}$ and $z$, denoted
$R(L_{X}, z)$. This can be directly related to the fraction of type 2
AGN, $f_2$, at any $(L_{X},z)$ by $f_2 = R/(1+R)$. As is
commonplace, we define type 2 AGN to be those that have absorbing
column densities $N_{\mathrm{H}} \geq 10^{22}$~cm$^{-2}$ in the X-ray
band. If the unified model is correct then $f_2$ is approximately
equal to the covering factor of the $\geq 10^{22}$~cm$^{-2}$ gas around
the AGN. However, the exact distribution of this material is
unknown. That is, the fraction of AGN observed
to have the various obscuring columns, the \nh\ distribution, is unknown
except for local bright Seyfert 2s \citep[e.g.,][]{rms99}. Under the
unified model, all these columns of gas exist close to the black hole,
with the observed \nh\ distribution giving the relative covering
fraction of each. This distribution may also be a function of
$L_{X}$ and $z$. Therefore, constructing a CXRB synthesis model
reduces to producing a model for how the X-ray absorption may vary
with luminosity and redshift.

Two different assumptions on the \nh\ distribution were used by
\citet{bem06} and we adopt the same distributions here. Ten values of
\nh\ were considered: $\log
N_{\mathrm{H}}=20,20.5,\ldots,24.0,24.5$\footnote{There do exist
examples of completely Compton thick AGN with estimated columns of
$\log N_{\mathrm{H}} \geq 25$ (e.g., NGC~1068; \citealt{matt97}). The
X-ray emission from these sources is so reduced that they are not an
important contributor to the CXRB, but they could potentially add to
the mid-IR AGN number counts and luminosity functions. However, test
calculations showed that the inclusion of such sources did not change
our \nh-averaged SEDs in the mid-IR from which we make our
predictions. Thus, objects with columns of $\log N_{\mathrm{H}} \geq
25$ can be safely excluded from our calculations. See Sect.~\ref{sub:seds} for
details on the construction of the \nh-averaged SEDs.}. In the
`simple \nh\ distribution', any type 1 AGN has an equal probability
$p$ of being absorbed by a column of $\log N_{\mathrm{H}} = 20$ or
$\log N_{\mathrm{H}} = 21.5$. Likewise, a type 2 AGN had an equal
chance of being absorbed by a column of $\log N_{\mathrm{H}} = 22$ or
$\log N_{\mathrm{H}} = 24.5$:
\begin{equation}
N_{\mathrm{H}} = \left\{ \begin{array}{lll}
                         20.0,\ldots,21.5 & & p=(1-f_2)/4.0\\
                         22.0,\ldots,24.5 & & p=f_2/6.0
			 \end{array}
                 \right .
\label{eq:simple}
\end{equation}
The second
assumed \nh\ distribution was the observed \citet{rms99} distribution for
type 2 AGN, and the simple one for type~1. \citet{rms99} found that in
a carefully selected sample of local Seyfert 2 galaxies, 75\% had
$\log N_{\mathrm{H}} \geq 23$ with half being Compton-thick ($\log
N_{\mathrm{H}} \geq 24$):
\begin{equation}
N_{\mathrm{H}} = \left\{ \begin{array}{lll}
                         20.0,\ldots,21.5 & & p=(1-f_2)/4.0\\
                         22.0,\ldots,23.5 & & p=f_2/8.0\\
			 24.0,24.5 & & p=f_2/4.0
			 \end{array}
                 \right .
\label{eq:risaliti}
\end{equation}

\citet{bem06} considered many different parameterizations for possible
evolutions of $R(L_{X}, z)$, and presented results for two cases that
could fit the observed shape of the CXRB and X-ray number counts (the
results were independent of the \nh\ distribution assumed). The first
evolution had an initial $z=0$, $L_{X}=10^{41.5}$~erg~s$^{-1}$ type
2/type 1 ratio of $R_0=4$, comparable to previous optical measurements
\citep{mr95}, and required only gradual redshift evolution:
\begin{equation}
f_2 = K (1+z)^{0.3} (\log L_{X})^{-4.8},
\label{eq:4to1}
\end{equation}
where $K$ is a constant defined by $R_0$. This evolution of the
absorbing gas around AGNs also provided an acceptable fit (reduced
$\chi^2 = 1.3$) to the type 1
fractions measured by \citet{bar05} at different luminosities and
redshifts. The second parameterization presented by \citet{bem06} was
inspired by the SDSS measurement of $R_0=1$ \citep{hao05}, and
required much more rapid redshift evolution in order to fit the CXRB:
\begin{equation}
f_2 = K (1+z)^{0.9} (\log L_{X})^{-1.3}.
\label{eq:1to1}
\end{equation}
This model did a poor job fitting the Barger \etal\ type 1
fractions (reduced
$\chi^2 = 2.6$). In both these cases, the $z$ evolution was halted at $z=1$,
because there was no constraint on $f_2$ at higher redshifts. Also,
the increase in $f_2$ with $z$ may slow significantly at $z \sim 1$ if
the gas obscuring the AGN is connected to the cosmic SFR density.

The CXRB spectrum and X-ray number-counts can also be fit with a model
where the type 2/type 1 AGN ratio does not evolve with redshift
\citep[e.g.,][]{tu05}. This scenario should be considered as a
null-hypothesis in our investigation. One of the unevolving
parameterizations that \citet{bem06} presented that fits the CXRB
spectrum is
\begin{equation}
f_2=K \cos^2 \left ({\log L_{X} - 41.5} \over 9.7 \right ),
\label{eq:noevol}
\end{equation}
where $R_0=4$. This model does have a modest luminosity dependence,
but this will not affect the comparison between the $z$-dependent
evolutions above.

Figure~\ref{fig:contours} shows contours of $R(L_X,z)$ for the three
evolutions presented here. The different models have very different
predictions for the ratio of obscured to unobscured AGN particularly at
$0.5 \leq z \leq 1$. The rapidly evolving model (Eq.~\ref{eq:1to1})
predicts ratios of 4:1 to 5:1 for quasars at $z \sim 1$, as opposed to
the 2:1 to 1:1 ratios predicted by the other two evolutions. At lower
luminosities and redshifts, the slower evolution model
(Eq.~\ref{eq:4to1}) predicts a larger value of $R$ than the other
possibilities. This paper will explore how these differences manifest
themselves in the IR AGN number counts and LFs. 

\section{Calculation of AGN Spectral Energy Distributions}
\label{sect:calcs}

\subsection{General Methodology}
\label{sub:methods}
If the gas and dust causing the X-ray absorption are also responsible
for the majority of the IR emission from an AGN, then it may be
possible to discriminate among the three evolutions by mid-IR
measurements. In this case, the IR emission for a given X-ray \nh\ can
be predicted by a photoionization model, and since the evolution of
\nh\ is predicted by the CXRB models, the subsequent evolution in
different IR bands can be predicted. The photoionization models also
allow the ability to vary parameters such as the location, density
distribution and dust content of the absorbing gas, which may result
in constraints to some of these physical properties.

We employ the photoionization code \cloudy\ v.\ 05.07.06 \citep{fer98}
to compute the IR emission of the absorbing gas. Although the last
decade has seen significant work on the exact radiative transfer
problem of IR photons through a dust torus
\citep{gd94,err95,gra97,dv05,ffh06}, these techniques do not provide an easy
correspondence to the X-ray absorbing gas. Furthermore, they ignore
gas-grain interactions such as gas-heating by photoelectric emission
and grain charging by electron capture \citep{wd01}. These effects are
included in the latest versions of \cloudy\ which also uses a new,
more physical grain model where the size distributions of both the
silicate and graphite components as well as PAHs are each resolved
into 10 size bins. The dust module in \cloudy\ also takes into account
the non-equilibrium temperature spiking process that affects the very
small grains \citep{pur76,sell84,vh04}. The computed SEDs therefore
include all the detailed dust physics of the latest torus models, but
also allow a connection to the X-ray properties. 

The disadvantage to this technique is that it requires a
simplification of the actual IR emission region which may be quite
complex \citep[e.g.,][]{nie02,dv05}. Furthermore, there are many
instances of local AGN whose X-ray absorbing gas seems to be mostly
dust-free \citep{wm02}, or where only a fraction of the total column
is dusty \citep{bwm03}. In addition, the radiative transfer techniques
employed in \cloudy\ are less sophisticated than those used in the
multi-dimensional torus models \citep{fer03}. However, our goal is not
to reproduce the SEDs of individual objects, but to make predictions
for the average properties of an ensemble of AGN at a given $L_{X}$
and $z$. Since even in local AGN, there is a wide range of IR
properties \citep{weed05,buch06}, it is plausible that no systematic
errors will result from the simplicity of the model calculations. This
expectation is tested against real data in Sect.~\ref{sect:tests}.

\subsection{\cloudy\ Model Setup and Assumptions}
\label{sub:assume}
Each \cloudy\ model has a simple setup: a constant AGN spectrum with a
2--10~\kev\ luminosity $L_X$ strikes a cloud with an inner radius
$r$~pc from the continuum source. The gas has abundances similar to
the Orion nebula (using the `abundances Orion' command in \cloudy),
and has a uniform hydrogen density of $10^4$~cm$^{-3}$, which is
typical for a molecular cloud. The calculation proceeded through the
cloud until the X-ray measured column density reached a value \nh\
(utilizing the `stop effective column density' command\footnote{We
confirmed that the X-ray hardness ratio derived from the \cloudy\
models using this method agrees with ones calculated with a
\texttt{wabs*pow} model in XSPEC.} ). The ionizing spectrum was
defined using the `agn' continuum in \cloudy\ with an X-ray
photon-index of 1.9, a big blue bump temperature of $1.4\times 10^5$~K
(appropriate for $10^7$~M$_{\odot}$ black hole accreting at one-tenth
of its Eddington rate), and an $\alpha_{\mathrm{ox}}=-1.4$. While this
is a common value of $\alpha_{\mathrm{ox}}$ over a wide range of
luminosity and redshift, \citet{stef06} showed that this parameter is
anti-correlated with the UV luminosity of the AGN. There is
significant scatter in this relation, however, and to date only
optically selected, very high luminosity, $z > 4$ quasars have been
found to typically exhibit significantly lower values of
$\alpha_{\mathrm{ox}}$ \citep[e.g.,][]{shem06}. Unfortunately, it is
unknown if there is any relation between $\alpha_{\mathrm{ox}}$ and
the infrared properties of AGN.

\cloudy\ has two stored grain (graphite and silicate)
distributions. One, denoted `Orion', is based on the flatter reddening
observed in star-forming regions due to a lack of small grains. The
other, called `ISM', reproduces the standard Galactic reddening
law. Since it has been argued that small dust grains close to an AGN
will be preferentially destroyed \citep{mmo01,wm02}, we employ the
Orion grains in our models; however, two grids with the ISM grains
were also computed to check the sensitivity of the results to this
assumption. PAHs were included in all calculations.

Finally, consistent with the unified model, we set the covering factor
of the illuminated gas and dust equal to $f_2$ when $N_{\mathrm{H}}
\geq 10^{22}$~cm$^{-2}$ or $1-f_2$ otherwise. This assumption may run
into trouble at high luminosities if, as has been proposed
\citep[e.g.,][]{san89}, quasars evolve over time from being completely
buried to unobscured. In this model, the covering factor of the
obscuring gas and dust need not be related to the fraction of type 2
sources at any ($L_X$,$z$) pair. However, for the time being we will
continue to assume this is the case and that the unified model is
valid over all luminosities. Sect.~\ref{sub:constantcovering} tests
the validity of this assumption.

Since the covering factor is variable between $z=0$ and $1$ when using
the evolutions described by equations~\ref{eq:4to1} and~\ref{eq:1to1},
\cloudy\ models also had to be calculated as a function of $z$ for
each $L_X$ and \nh. For completeness, the cosmic microwave background
with the appropriate temperature and intensity for the particular $z$
was added to the illuminating spectrum using the `cmb' command.

\subsection{Choosing the Inner Radius of the Absorbing Gas}
\label{sub:inner}
The minimum distance from the radiation source that a dust grain can
typically survive is the sublimation radius, which is a function of
both the grain size and composition as well as the luminosity of the
AGN \citep{bar87,ld93}. For a bolometric AGN luminosity of
$10^{46}$~erg~s$^{-1}$, \citet{ld93} show that subliming grains
exist from $\sim 0.2$~pc (for the largest grains) all the way to $\sim
4$~pc (for the smallest grains) from the central engine. Since the
\cloudy\ models resolve the grains in both size and composition, we are
unable to pick an inner radius for the absorbing material that will
self-consistently account for subliming grains over all grain
radii. For example, if we chose an inner radius of $\sim 4$~pc so that
only the smallest grains are subliming then we are neglecting the
near-IR emission from the larger grains that can exist much closer to
the AGN. In contrast, if we pick a radius of $\sim 0.2$~pc, the
sublimation radius for the largest grains, then the smaller grains
would be above their sublimation temperatures and radiating at
inappropriately small wavelengths. Self-consistently accounting for
grain destruction is a complicated procedure especially if it occurs
only over a fraction of the cloud depth. Therefore, as a compromise
between the two extremes, we have selected the inner radius of the
absorbing cloud to be $r=1$~pc. For the Orion (ISM) grain
distributions, sublimation temperatures for the smallest grains begin
to be reached at $\log L_X =
43.5$ ($43.25$) and no attempt is made to correct the results
for grains radiating above their sublimation temperature. This will
have the largest effect on SEDs computed at the highest luminosities
($\log L_{X} \gtrsim 46$) at which a significant portion of the hot
dust emission is shifted to shorter wavelengths than
observed. These high luminosity objects are rare, and as our
predictions depend on the AGN luminosity function, this issue will
have only a small effect on our predictions from $z=0$--$1$. Moreover,
Sect.~\ref{sect:tests} shows that when averaged over the \nh\
distribution, the computed SEDs have very similar IR properties to the
ensemble of observed AGN.

Choosing a fixed inner radius for the attenuating cloud is a different
but equally valid technique as that
employed by recent models of dust tori, which select the
inner radius for a single grain size and change it with AGN luminosity
\citep[e.g.,][]{ffh06}. This latter strategy may be more appropriate
for fitting individual SEDs, but we are most interested in determining
the properties of average AGNs that are being selected in the deep
\chandra\ and \spitzer\ surveys. Thus, it is more
appropriate to think of the inner radius of $r=1$~pc as an average distance or as
a radius where there is a significant density enhancement in an
average AGN.

In this context, we also calculated models with the inner radius at
$r=10$~pc. As mentioned in \S~\ref{sect:intro}, there is significant
observational evidence that dust emission over a range of radii is
required to fit the observed SEDs of many AGN. In addition, if the
obscuring material is connected to a galactic-scale phenomenon such as
a starburst ring, then a significant density enhancement at distances
of $\sim 10$~pc may be common. It is therefore interesting to check if
the ensemble of AGNs shows evidence for enhanced density at these
distances. Clearly, dust emission from $\sim 10$~pc may be too cool to
account for some of the observed near-IR AGN properties (sublimation
temperatures were reached at $\log L_{X} = 45.5$ for the Orion
grains), but it may be important in the mid-IR.

\subsection{The Model Grids}
\label{sub:grids}
Our interest is in predicting the mid-IR properties of AGN as a
function of the 2--10~\kev\ X-ray luminosity $L_X$ and $z$ using the
fits to the CXRB as constraints. Thus, we iterate from $\log L_X=41.5$
to $48$ (in steps of 0.25), and $z=0$ to $5$ (in steps of 0.05) as we
did in constructing the CXRB synthesis model \citep{bem06}. For each
($\log L_X$, $z$) pair we compute using the evolution
equations~\ref{eq:4to1},~\ref{eq:1to1} or~\ref{eq:noevol} the value of
$f_2$, and therefore the covering factor of absorbing gas around the
black hole (if the unified model is correct). A \cloudy\ model is then
computed for each $\log N_{\mathrm{H}} = 20.0, 20.5, \ldots, 24.0,
24.5$~cm$^{-2}$. In practice, because the redshift evolution is only
constrained from $z=0$ to $1$, we assume that there is no redshift
evolution in $f_2$ when $z > 1$ (see also
\S~\ref{sect:cxrbreview}). Therefore, the photoionization models were
calculated only up to $z=1$, and the $z=1$ models are then used at $z
> 1$. This resulted in 5670 individual \cloudy\ models for the two
redshift-dependent evolutions (eqs.~\ref{eq:4to1} and~\ref{eq:1to1})
and only 270 models for the unevolving torus model
(eq.~\ref{eq:noevol}).

Grids were computed for all three evolutions with the inner radius of
the absorbing cloud at $r=1$~pc and at $r=10$~pc. For ease of
reference, we will now refer to results from individual grids using a
compact label based on the initial $R_0$ and $r$ of the grid (e.g.,
4:1/1pc, 1:1/10pc, or nozevol/1pc).

\subsection{Constructing the Final Spectral Energy Distributions}
\label{sub:seds}
At this point, within every grid, there are 10 \cloudy\ models at each
($L_X$,$z$) pair for the different \nh\ values. The photoionization
models provide several different spectra for each computation. The
ones of most interest for these purposes include the sum of the
transmitted continuum and outward directed diffuse emission from the
illuminated material (corresponding to observing the AGN through the cloud),
and the reflected emission from the inner surface of the
cloud. However, under the unified model, since all columns exist
around the AGN, we would observe reflected emission from all \nh\
including the Compton-thick columns. Evidence for this can be found in
the X-ray spectra of many Seyfert 1s where narrow \fe\ lines are
observed from optically-thick material out of the
line-of-sight \citep[e.g.,][]{yp04}. Thus, to construct the SED for
each \nh, a weighted average of the reflected emission from all 10
\nh\ models is added to the transmitted+outward diffuse spectrum for
that particular \nh. The weights in the average followed the chosen
\nh\ distribution, so that for the `simple \nh\ distribution' the
reflected spectrum for all the $N_{\mathrm{H}} \geq 10^{22}$~cm$^{-2}$
models had a weight of $f_{2}/6$ (eqn.~\ref{eq:simple}). In this way, we calculate a spectrum
for each \nh\ that corresponds to observing the AGN through that
column but also takes into account the reflected emission from the
other columns that lie out of the line of sight. We refer to these
SEDs as the `unified SEDs'.

The last step before calculating specific predictions for the mid-IR
wavebands is to average these unified SEDs together to find an average
AGN SED at this luminosity and redshift. This is the same procedure as
is used when constructing CXRB synthesis models
\citep[e.g.,][]{gill99,gill01}. Two different \nh\ distributions
(described in Sect.~\ref{sect:cxrbreview}) are used to generate this
final SED. Figure~\ref{fig:sedexample} shows examples from the 4:1/1pc
and 4:1/10pc grids when $\log L_X =43$ and $z=0.45$ (the spectra are
plotted in the rest frame). At this luminosity and redshift,
eqn.~\ref{eq:4to1} gives a type 2 AGN fraction of 0.754 ($R \approx
3$), so the \nh-averaged spectrum will be dominated by the `unified
SEDs' with \nh$\geq 10^{22}$~cm$^{-2}$. The Risaliti \etal\
distribution, with half of the type 2 AGN being Compton-thick, only
makes a significant difference to the average SED at wavelengths
$>30$~\micron. The X-ray part of the averaged SED is only moderately
absorbed as is required to fit the CXRB (see below).

An interesting property of the \nh-averaged SED shown in the top panel
of Fig.~\ref{fig:sedexample} is the strength of the silicate emission
at $\sim 10$\micron\ even though at large values of \nh\ silicate
absorption is commonly observed. This emission, as well as the
majority of the other emission lines, is a result of including the
reflected emission from the hot inner walls of the absorbing
cloud. The fact that silicate emission is predicted even for spectra
which are dominated by type 2 AGN is interesting given that the
majority of AGN with typical Seyfert~2 levels of \nh\ show silicate
absorption \citep{shi06}, although some exceptional cases do show
emission \citep{sturm06}. As our main focus is on the overall
continuum properties, the behavior of this feature, in addition to the
emission lines, will not significantly influence the predictions. We
do note, however, that the \nh-averaged SED from the $r=10$~pc model
(lower panel in Fig.~\ref{fig:sedexample}) exhibits significantly
weaker silicate emission than in the case where the inner radius of
the attenuating material lies at 1~pc.

\section{Properties of the Model SEDs}
\label{sect:tests}
Given the simplifying assumptions made in these calculations, it is
important to verify that the \nh-averaged SEDs are fair
representations of reality, particularly in the mid-IR where the
predictions of the number-counts and LFs will be made. Our goal is not
to fit the SEDs of individual objects, but to compare various properties
of the ensemble of models to the large samples of AGN being studied in
the deep, multi-wavelength surveys. This is valid because we will only
be making predictions (cumulative number count distributions and
luminosity functions) that will be compared against data that are
averaged over many individual objects.

The photoionization models greatly ease the ability to make the
comparisons to data, as we are able to calculate the observed X-ray
flux (in addition to the IR fluxes) in multiple bands. In this way we
can make true multiwavelength comparisons to data.

The `simple \nh\ distribution' (eqn.~\ref{eq:simple}) was used for the
models presented throughout this section. Employing the \citet{rms99}
distribution (eqn.~\ref{eq:risaliti}) only makes a very minor
difference to observables like colors, so these results are not
presented here.

In this section and throughout the paper, model IR fluxes were
computed using the predicted flux at the specific wavelength of
interest (e.g., 3.6\micron, 8\micron, etc.). \spitzer\ of course
measures an average flux over a passband. Comparing fluxes computed at
one wavelength with those averaged over the \spitzer\ passband yielded
differences of a few percent. Given the level of the
systematic assumptions made in the calculations, this difference in
predicted fluxes is negligible. For this reason we also do not
consider the wavelength dependent efficiency of the \spitzer\
passbands.

\subsection{Mid-IR to X-ray Flux and Luminosity Ratio}
\label{sub:irtoxray}
If the attenuating gas and dust around an AGN was strictly responsible
for the thermal IR emission, then a correlation might be expected
between the relative strength of the IR flux and the absorbing column
density. That is, the most obscured sources should produce the largest
mid-IR to X-ray flux ratio. However, mid-IR observations of X-ray
selected AGN have shown no correlation at all between the X-ray
hardness ratio (a proxy for the absorbing column \nh\ since more
heavily absorbed objects will appear harder) and the strength of the
IR flux relative to the hard X-ray \citep{lutz04,rig04}.

In Figure~\ref{fig:nhhard} we test the existence of this expected correlation with the `unified
SEDs' computed from the \cloudy\ models. Panel (a) plots the
$f_{24\micron}/f_{\mathrm{2-8\ keV}}$ ratio as a function of the
$f_{\mathrm{2-8\ keV}}/f_{\mathrm{0.5-2\ keV}}$ ratio for all the
`unified SEDs' with $z < 3$ in the 4:1/1pc grid (the results are
not significantly different using other evolutions). Panel (b) shows
the same data except only includes the results for models with $\log
L_{X} < 43$. The black region shows where models with $\log N_{\mathrm{H}} <
23$ fall, the red area plots models with $23 \leq \log N_{\mathrm{H}} < 24$,
and the blue region denotes models with $\log N_{\mathrm{H}} \geq 24$. The
green squares are combined \chandra\ and \spitzer\ data on AGN in the
\chandra\ Deep Field South (CDFS; \citealt{rig04}). 

The first important result to note is that only objects
with columns $\log N_{\mathrm{H}} \geq 24$ produce the largest relative
24\micron\ fluxes. This is in agreement with the results of
\citet{rig04} who employed local observed SED templates to predict the
$f_{24\micron}/f_{\mathrm{2-8\ keV}}$ ratio.  Panel (b) of
Fig.~\ref{fig:nhhard} shows that the lowest
$f_{24\micron}/f_{\mathrm{2-8\ keV}}$ ratios are found from objects
with $\log L_{X} > 43$ if the inner radius of the absorber is at 1~pc.
However, if $r=10$~pc then the absorber is less ionized for a given
$\log L_X$, and larger hardness ratios are possible, which enable
these models to fill
much more of the observed parameter space (panels (c) and
(d)). Interestingly, panel (d) shows that when the inner radius of the
attenuating material is at 10~pc, the lowest
$f_{24\micron}/f_{\mathrm{2-8\ keV}}$ ratios are found from objects
with $\log L_{X} < 43$, the opposite of what was found when
$r=1$~pc. This can be understood by using Fig.~\ref{fig:sedexample} as
a guide. The $r=1$~pc models produce significant hot dust emission
that peaks at wavelengths $< 24$\micron, while the 10~pc models have
much cooler dust emission peaking at 24\micron\ or even longer (the
2--8~\kev\ flux is basically identical between the two
cases). Therefore, as the spectra are redshifted, the hot dust
emission in the $r=1$~pc models moves through the 24\micron\ band and
keeps the mid-IR/X-ray ratio high. In contrast, the flux at 24\micron\
would be lower at the same $L_{X}$ and $z$ for the $r=10$~pc models
since the peak mid-IR emission would be redshifted out of this
band. The CDFS survey, being a pencil-beam deep field, is dominated by
AGN with $\log L_X < 44$ \citep{zhe04}. This would seem to indicate
that the observed-frame 24\micron\ emission may originate, on average, from material
a few pc removed from the AGN.

It is also interesting to note the few objects with enhanced
$f_{24\micron}/f_{\mathrm{2-8\ keV}}$ ratios that are found at very soft
hardness ratios. None of the computed `unified SEDs' were able to
account for these sources. They may have increased 24\micron\ flux due
to star-formation heating. However, given the significant number of
assumptions made in these calculations on ionizing spectrum, chemical
composition, geometry, etc., it is more important that the models
account for the broad properties of the ensemble.

Figure~\ref{fig:hrdratioavg} plots the same quantities as
Fig.~\ref{fig:nhhard}(a) except the measurements are made from
\nh-weighted spectra. As alluded to in the previous section the X-ray
hardness ratios of these SEDs do not span a wide range. This is
necessary to fit the CXRB which is well described by a $\Gamma=1.4$
power-law between 1 and 20~\kev\ \citep{kush02,lumb02,dm04}. In
Fig.~\ref{fig:hrdratioavg} we denote the $f_{\mathrm{2-8\
keV}}/f_{\mathrm{0.5-2\ keV}}$ ratio for such a power-law with the
vertical dotted line. The red region denotes the
$f_{24\micron}/f_{\mathrm{2-8\ keV}}$ ratio measured from models with
$z < 1$ and $\log L_{X} < 44$. The recent deep X-ray surveys all show
that these AGN dominate the production of the 2--10~\kev\ CXRB
\citep{ueda03,bar05,laf05}. Indeed, the red area is bisected by
the hardness-ratio appropriate for producing the correct shape of
the CXRB.

Fig.~\ref{fig:nhhard} shows that the model SEDs, which were
constructed based on the assumptions of the unified model, can fully
account for the distribution of data points from the CDFS. In fact, no
correlation between $f_{24\micron}/f_{\mathrm{2-8\ keV}}$ and the
hardness ratio is predicted by these models. This is, in part, due to
the poor translation from hardness ratio to absorbing \nh\ column for
surveys that span a large range in redshift. For a given \nh, the
observed hardness ratio will decrease with increasing redshift. To
correct for this problem, \citet{rig06} and \citet{lutz04} plotted the
ratio of the absorption-corrected 2--10~\kev\ luminosity to the
6\micron\ luminosity versus the inferred \nh\ for AGN with known or
estimated redshifts. These authors again found no correlation of the
X-ray-to-mid-IR luminosity with absorbing column at $z=0$ or at
$z=1$. Figure~\ref{fig:rigby06} shows the predicted $L_{X}/\nu
L_{\nu}$(5.7\micron) versus $\log N_{\mathrm{H}}$ plot for the unified
SEDs resulting from the 4:1/1pc and 4:1/10pc grids. As the ratios are
constructed using rest-frame quantities, only one redshift is plotted
($z=0.7$). The predicted ratios are color-coded based on X-ray
luminosity with black triangles corresponding to $46 < \log L_{X} \leq
48$, red triangles denoting $44 < \log L_{X} \leq 46$, and blue
triangles showing the results for $41.5 \leq \log L_{X} \leq 44$. The
green squares and cyan stars plot the ratios observed by \citet{rig06}
and \citet{lutz04}, respectively.

Both the 1~pc and 10~pc models cover roughly the same region in the
$L_X$ to $\nu L_{\nu}$(5.7\micron) ratio, but can only explain about
the upper half of the observed data points. Evidently, these unified
SEDs do not always produce enough 5.7--6\micron\ emission to account
for the observations. Star-formation is a possible culprit, but
\citet{lutz04} correct for this by decomposing their spectra using the
PAH emission as a guide to the star-formation strength. This effect
may also be a result of omitting the host galaxy from the calculated
SEDs, which will become more important at low AGN luminosities (we
return to this point in Sect.~\ref{sub:fluxes}). \citet{lutz04} do
suggest this may be a possible source of error in their measured
6\micron\ fluxes, but it is unlikely to account for the roughly factor
of 10 disagreement for some of the points. Perhaps the disagreement is
a result of missing hot dust emission from within 1~pc. Replotting the
figure with results using the ISM grain distribution that includes
smaller (and hence hotter) grains results in only a very small
decrease in the $L_X$ to $\nu L_{\nu}$(5.7\micron) ratio. This result
indicates that more hot dust emission may be helpful but is again
unlikely to solely account for the large discrepancies implied by some
of the data points. It seems more likely that the disagreement is due
to a combination of all three effects, as well as possible measurement
errors and the failure of one of our fundamental assumptions in the
geometry of the absorber. A full investigation of this problem is
beyond the scope of this paper and is left for future work.

Finally, it is interesting that in the 4:1/1pc models, the largest
$L_X$ to $\nu L_{\nu}$(5.7\micron) ratios are found at the highest
$L_X$, but, for the 4:1/10pc models, they are found at the lowest
$L_X$. Indeed, when $r=10$~pc, roughly the same spread of $L_X/\nu
L_{\nu}$(5.7\micron) is found for $\log L_{X} > 46$ and $\log L_{X}
\leq 44$. When the absorbing material is closer to the ionizing
source, it is easier to produce hot dust emission at low X-ray
luminosities, so as $L_{X}$ increases, there is a slower corresponding
increase in $\nu L_{\nu}$(5.7\micron). At larger distances, it is more
difficult to produce hot dust and, for low $L_X$, increasing $L_X$
also results in little response at $\nu
L_{\nu}$(5.7\micron). The sources plotted by \citet{rig06} have $\log
L_{X} \lesssim 44.7$, and are thus consistent with either value of $r$.

\subsection{Infrared Colors}
\label{sub:colors}
Another important check on our calculation strategy is the
distribution of the observed-frame IR colors. In
Figs.~\ref{fig:lacycolor} and~\ref{fig:lacycolor2} color-color
diagrams are constructed from the \nh-weighted SEDs and compared with
\spitzer\ data presented by \citet{lacy04}. In the first case, which
examines only IRAC colors, the vast majority of the models fall within
the selection region determined by
\citet{lacy04}. Fig.~\ref{fig:lacycolor2} compares a MIPS color to an
IRAC one.  In both cases, although there are a few data points that are
significantly redder or bluer than the predicted colors, most of the
\nh-averaged SEDs have colors that are very similar to the observed
data.

The curves traced out by the models in both the color-color plots are
the changes caused by the IR slope (which depends on $L_X$) being
redshifted through the observed bands. For example, the `tail' at
constant $f_{\mathrm{5.7\micron}}/f_{\mathrm{3.6\micron}}$ in the
1:1/10pc panel of Fig.~\ref{fig:lacycolor2} is caused by AGN with
$\log L_{X} < 43$ over all $z$. At low $L_{X}$ and an inner radius of
10~pc, there is little hot dust emission, so the SED does not begin to
rise until $\sim 8$\micron\ (Fig.~\ref{fig:sedexample}). The
5.7\micron-to-3.6\micron\ slope is therefore nearly unchanged over the
$z$ range considered here, but
$f_{\mathrm{24\micron}}/f_{\mathrm{5.7\micron}}$ falls with redshift
as less emission gets redshifted into the 24\micron\ band.

Another useful way of comparing predicted to observed mid-IR colors is
the histogram of $\nu f_{\nu}(24\micron)/\nu
f_{\nu}(8\micron)$. \citet{brand06} found that the vast majority of
X-ray sources in the XBo\"{o}tes survey had $\nu
f_{\nu}(24\micron)/\nu f_{\nu}(8\micron) \approx 0$. In
Figure~\ref{fig:brandcolor}, histograms of this ratio are presented
from the \nh-averaged SEDs for both the nozevol/1pc and nozevol/10pc
grids (black lines), as well as the 1:1/1pc and 1:1/10pc grids (red lines). To more easily compare with the
\citet{brand06} result, the histograms only included SEDs with
$f_{\mathrm{0.5-7\ keV}} > 7.8\times 10^{-15}$~erg~cm$^{-2}$~s$^{-1}$,
the flux limit of the XBo\"{o}tes survey, and $f_{\nu}(24\micron) > 0.3$~mJy. In contrast to a strong peak
at $\nu f_{\nu}(24\micron)/\nu f_{\nu}(8\micron) \approx 0$, the
models computed with $r= 1$~pc tend to have bluer mid-IR colors with
the peak at $\nu f_{\nu}(24\micron)/\nu f_{\nu}(8\micron) \approx
-0.6$. The situation improves slightly when $r=10$~pc. In this case,
the histogram is much flatter with only a slight enhancement of
sources with negative values of $\nu f_{\nu}(24\micron)/\nu
f_{\nu}(8\micron)$. These
results indicate that models with the inner radius of the absorbing
material $\sim 1$~pc away from the AGN do not produce enough
24\micron\ emission to reproduce the mid-IR colors of an average
AGN. This flux could be added by heating by intense star-formation,
or, as indicated here, by emission from material at larger distances.
The 1:1 evolution models produce fewer bluer SEDs than the unevolving
AGN model. This is likely a result of the greater fraction of obscured
large-$L_X$ AGN at $z \sim 1$.

\subsection{Absolute Fluxes}
\label{sub:fluxes}
The final test to be performed here is comparing the absolute
observed-frame fluxes
predicted by the \nh-weighted SEDs with the available
data. Figure~\ref{fig:barmbyplot} plots the predicted 8\micron\ and
3.6\micron\ fluxes against the predicted 0.5--10~\kev\ fluxes. The
black crosses are from the 4:1/1pc models, while the blue
triangles plot the predictions from the 4:1/10pc models. Only
models with $z < 3$ were plotted on the Figure. To compare with these
predictions, the IR and X-ray data from sources in the Extended Groth
Strip (EGS) were taken from \citet{barm06} and are plotted as red squares.

Fig.~\ref{fig:barmbyplot} reveals an important limitation in our
\nh-weighted SEDs. As expected, there is a strong correlation between
the IR and X-ray fluxes in our model, and this does a good job
describing the distribution of data from the EGS at high X-ray
fluxes. At low X-ray fluxes, the predicted IR flux severely
underestimates the average measured fluxes, in particular at
3.6\micron. This is because we do not include any emission from the
host galaxy in our model SEDs. This will become important at the
lowest AGN luminosities, especially
at 3.6\micron\ where there will still be an important contribution
from the old stellar population of the galaxy. This problem does
not seem to impact the 8\micron\ fluxes as strongly, but we will not be
able to make accurate predictions at 3.6\micron\ or
4.5\micron. As seen in the following sections, this will not be a
significant hindrance in producing strong predictions at longer
wavelengths.

As evidence for this, Figure~\ref{fig:f2to10vsf24} plots the predicted
2--10~\kev\ fluxes from the $z<3$ models in the 4:1/1pc and
4:1/10pc grids against the 24\micron\ flux. The symbols are the
same as the previous plot except the red squares plot data from the
CDFS presented by \citet{ah06}. The locus of model predictions bisects
the data points indicating that the calculated 24\micron\ flux gives a
good description of an average AGN seen in the deep surveys. Also
plotted in the figure are regions where bright local AGN (`Piccionotti
AGN') and local starburst galaxies are expected to be found (see
\citealt{ah04,ah06}). The \nh-weighted SEDs predict fluxes that place
them within or on the lower boundary of the local AGN properties,
again indicating that the models are good descriptions of AGN over a
range of redshifts and luminosities.

In summary, the computed SEDs seem to adequately describe average AGN
behavior from the X-ray to the mid-IR, a property that has previously
only been possible using SED templates. These results also indicate
that the assumption of a constant $\alpha_{\mathrm{ox}}$ is not
resulting in a serious bias or error. Our computational approach has
the advantage of being able to vary the physical properties of the
absorbing gas to see which arrangement best describes the average
AGN. The successful tests presented in this section validate this
strategy and give confidence that the calculated number counts and LFs
will have significant predictive power.

\section{Predictions for bare AGNs}
\label{sect:agn}
This section presents the predicted cumulative number count
distributions and luminosity functions resulting from the final
\nh-averaged SEDs.  Recall that our goal is to use the
results of the deep hard X-ray surveys and the CXRB to predict the IR
properties. Therefore, a key ingredient in computing these integrated
quantities is the hard X-ray luminosity function (HXLF) and its
evolution with $z$ and $L_{X}$. As before \citep{bem06}, we employ
the luminosity-dependent density evolution HXLF of
\citet{ueda03}. This type of evolution has also been inferred by
\citet{has05} and \citet{laf05} using other X-ray surveys.

\subsection{Number Counts}
\label{sub:counts}
The expression used to calculate the cumulative number counts at
different IR wavelengths is the same as equation~3 from
\citet{bem06}. However, an important difference in the calculation is that the
integral over X-ray luminosity is begun at the minimum $L_{X}$ that
results in an IR flux $S$. In this way, for each IR wavelength of
interest, number count distributions are computed for each of the
three $R$ evolutions considered, and for models with
inner radii at 1 and 10~pc. Only the results with the simple \nh\
distribution are shown below, as there were only minor differences
when the \citet{rms99} distribution was assumed.

Figure~\ref{fig:counts} plots the predicted cumulative number counts
distributions at 5.7, 8, 24 and 70\micron. The black lines denote the
three evolutions computed with the inner radius of the absorbing gas
and dust at $r=1$~pc from the ionizing source, while the red lines
show the results when $r=10$~pc. Within each color, the three line
styles differentiate among the three evolutions of $R$. The data
points in the 5.7, 8 and 24\micron\ panels are the measured number
counts of X-ray selected AGN from the GOODS survey \citep{tre06}.

In the 5.7 and 8\micron\ panels there are two dashed green lines that
plot the computed number counts for the nozevol/ISM/1pc and
nozevol/ISM/10pc models. The smaller grains included in the ISM model
enhance the short wavelength IR emission, and therefore slightly
increase the number counts at a given flux for these two \spitzer\
bands. This difference is small enough that it will not affect the
arguments presented below.

It is clear from this figure that, apart from a small amount at
24\micron, there is little to no difference in the predicted number
count distribution for the three evolutions of the type 2/type 1
ratio. This is a result of integrating over $L_{X}$ and $z$ which will
tend to wash out any changes among the evolutionary models.
Differences in the number count distributions do arise at 70\micron, but they
are small in magnitude and it may be difficult for future surveys to
reduce the errorbars to small enough values to discriminate among
the different AGN evolutions.

The number count distributions do show a significant difference
between the ones based on calculations with $r=1$~pc (black lines) and
the models with $r=10$~pc (red lines). Before drawing conclusions it
is important to consider the consequences of both our methodology and
the observational selection effects on this plot. As mentioned above
(Sect.~\ref{sub:fluxes}), the omission of starlight from the host
galaxy results in the models underpredicting the near-IR flux at low
AGN fluxes. This will result in an underprediction of the number
counts at small IR fluxes, with the difference increasing both in
strength and in flux as we move to shorter wavelengths. The GOODS
survey covers a small angle on the sky and the data plotted here are
X-ray selected, so the survey will be missing both
bright, rare objects and heavily obscured AGN, both of which are
included in our calculations. Indeed, \citet{don05} and \citet{ah06}
both find that a significant fraction ($\sim 50$\%) of AGN have been
missed in the X-ray surveys, even at exposures $> 1$~Ms (see also
\citealt{bem06}). The addition of rare, bright objects will result in the models
overpredicting the number counts at the bright end. Compton-thick AGN,
while very faint in the 2--8~\kev\ X-ray band, can produce very
typical AGN fluxes in the IR \citep{ffh06}. The
inclusion of these sources in our models will result in an
overprediction compared to the data points at all fluxes, where the
amplitude of the error is related to the relative number of
Compton-thick AGN.

With the above points in mind, Fig.~\ref{fig:counts} indicates that
the $r=10$~pc models provide the best description of the observed
number counts at all wavelengths. This is perhaps not surprising at
observed-frame 24\micron\ as it is consistent with our earlier conclusions from
Fig.~\ref{fig:brandcolor} that the $r=1$~pc models do not on average
produce enough 24\micron\ flux. However, this conclusion at
observed-frame 8\micron\ seems at odds with Fig.~\ref{fig:barmbyplot}
which indicates that the $r=1$~pc models are a better
description of the 8\micron\ flux. With this interpretation, the
number counts should be observed to be underpredicted by our models at
low 8\micron\ flux, but this is only the case for the $r=10$~pc
models. The two plots can be brought into agreement only if a non-AGN
emission component begins to dominate the 8\micron\ flux below a
certain X-ray flux. Indeed, the 8\micron\ flux of the
EGS data is nearly constant below $f_{\mathrm{0.5-10\ keV}} \sim
10^{-14}$~erg~cm$^{-2}$~s$^{-1}$ (this is perhaps clearer in the
original Fig.~7 of \citealt{barm06}). Therefore, the $r=10$~pc models
are consistent with both the EGS 8\micron\ fluxes and the 8\micron\
GOODS number counts.

To summarize this section, we have found that the cumulative number
count distributions are very sensitive to the geometry of the
obscuring material of the AGN, and the GOODS data are best described
by the models where the attenuating material is $\sim 10$~pc from the
AGN. It will be interesting to compare these models to future AGN
counts at 70\micron.

\subsection{Luminosity Functions}
\label{sub:lfs}
\begin{figure}
\epsscale{1.0}
\plottwo{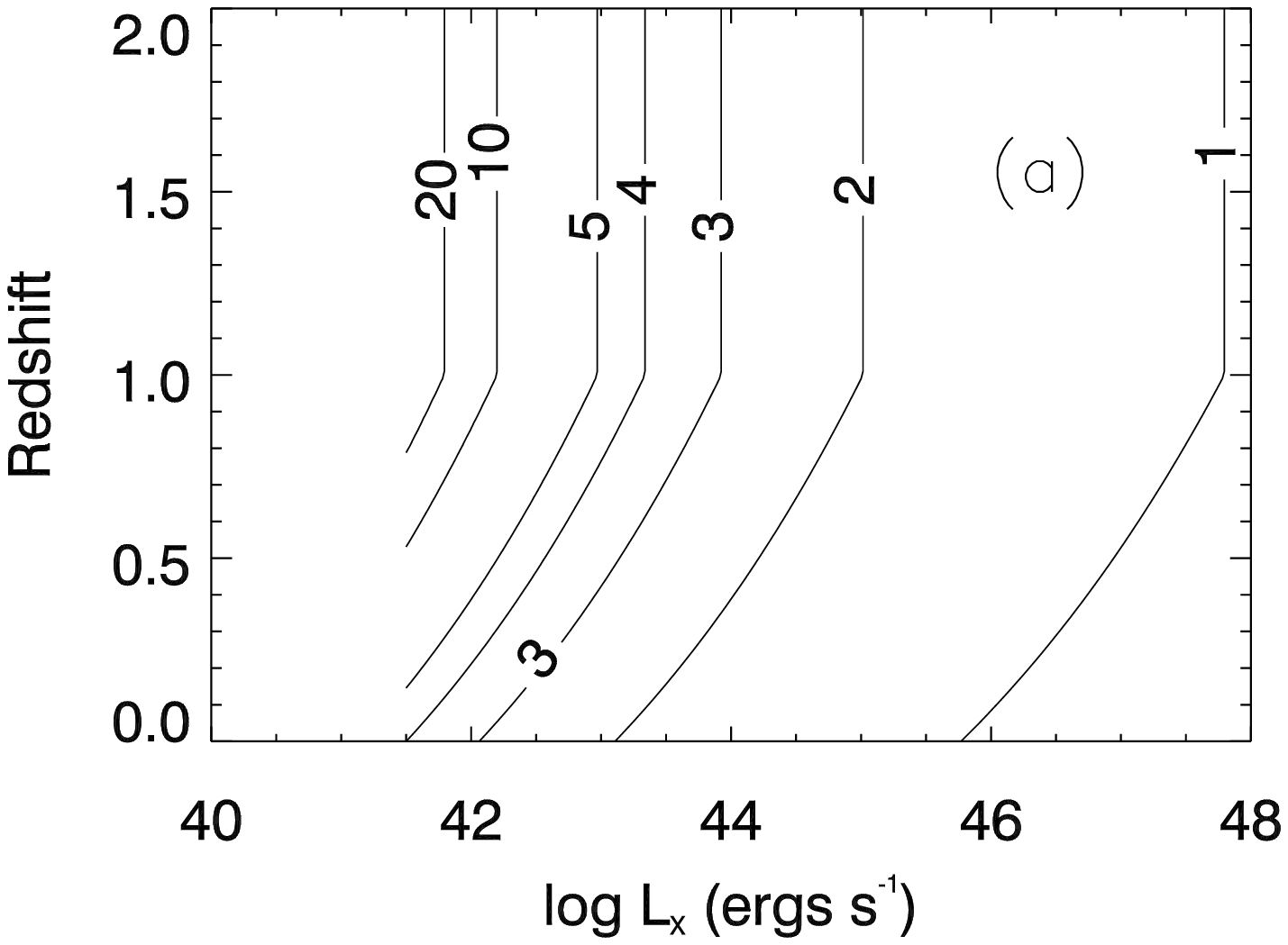}{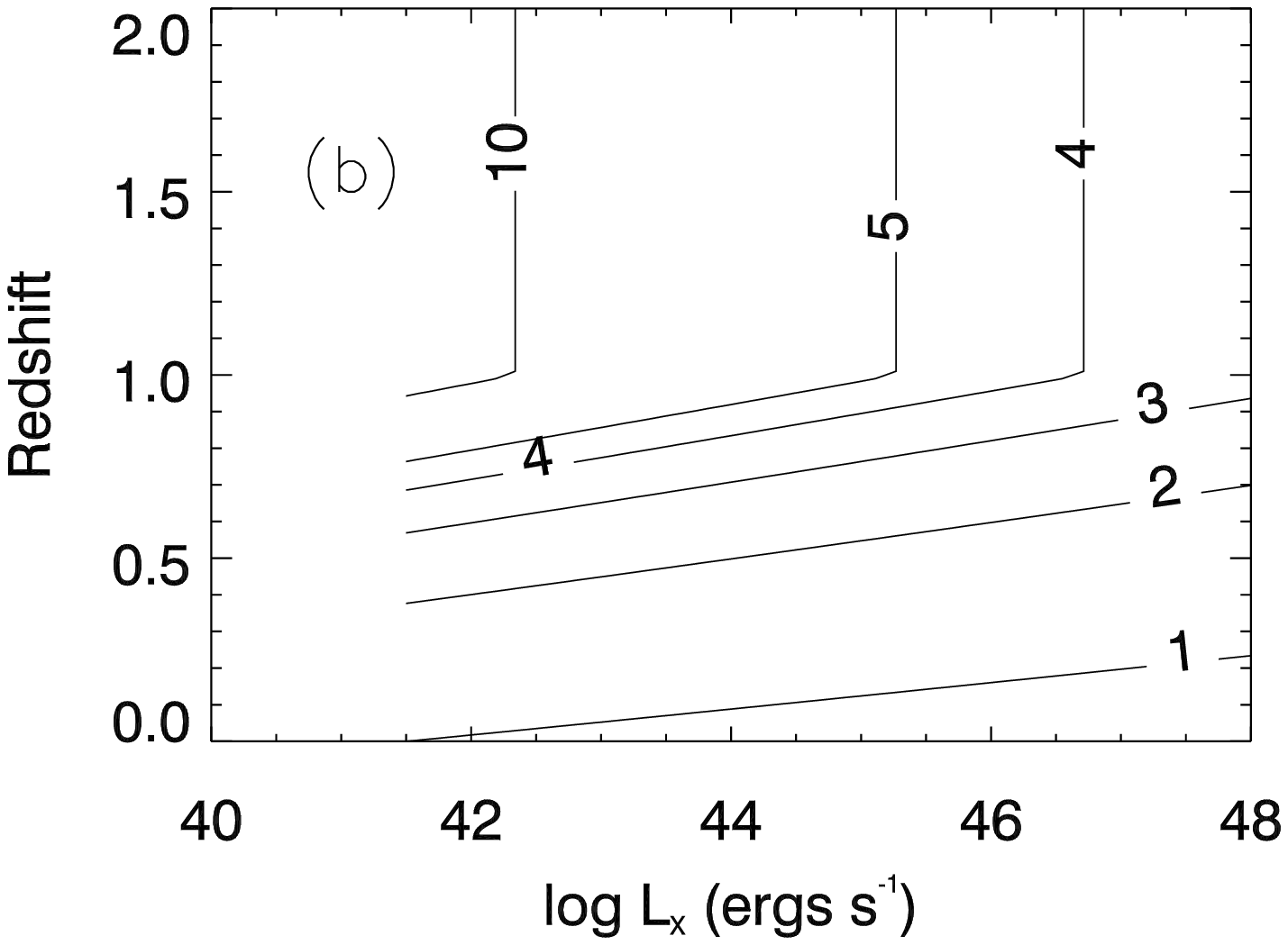}
\epsscale{0.5}
\plotone{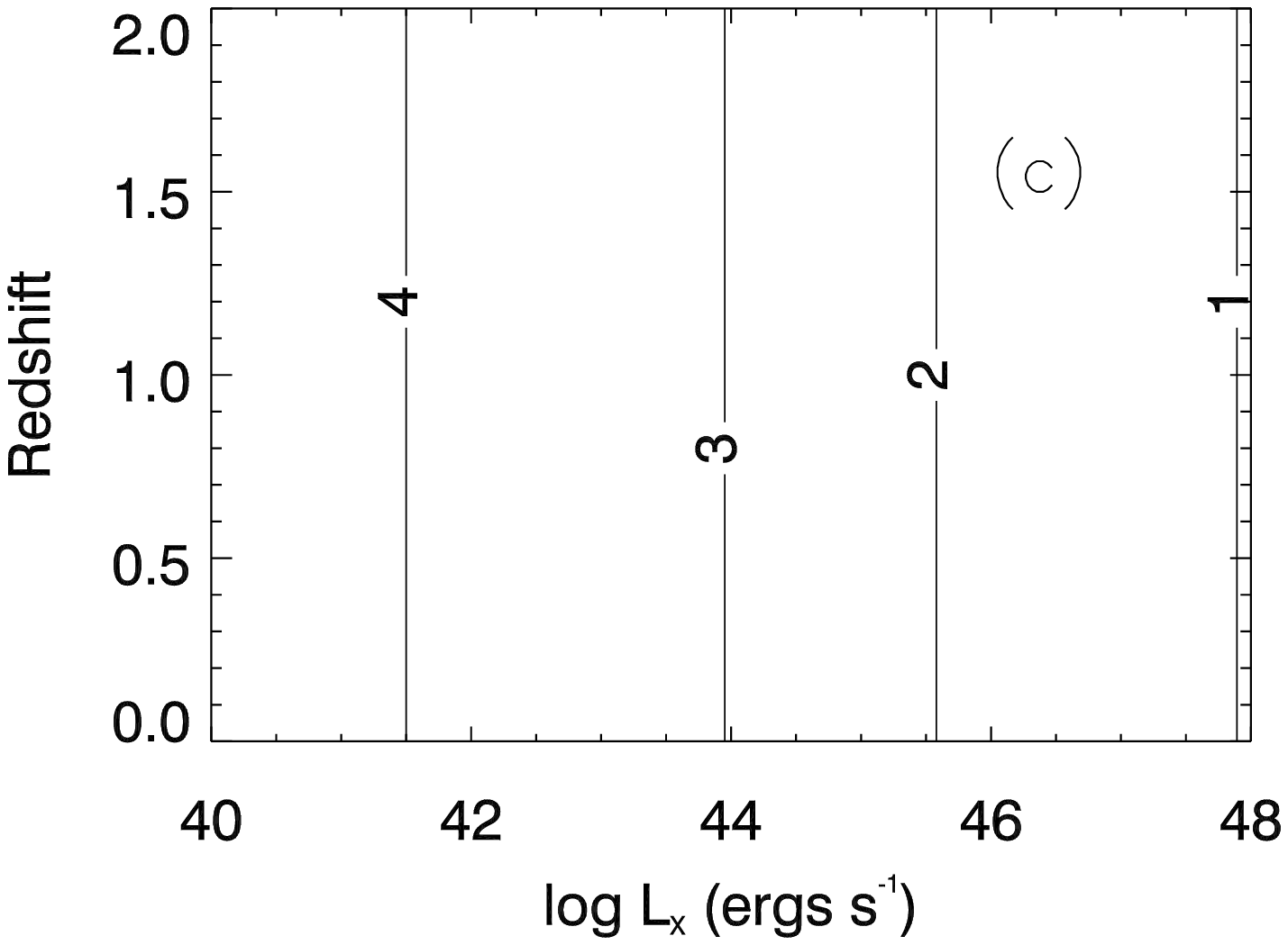}
\caption{(a) Contours of the AGN type 2/type 1 ratio ($R$) as a
function of $L_X$ and redshift for the evolution described by
eqn.~\ref{eq:4to1}. Adapted from \citet{bem06}. Recall that the
$z$-evolution is halted at $z=1$. (b) Same as (a), but shows contours
of $R$ for the more rapid evolution of eqn.~\ref{eq:1to1}. (c) As in
(a) and (b), but for a model with no redshift evolution (Eq.~\ref{eq:noevol}).}
\label{fig:contours}
\end{figure}

\begin{figure}
\begin{center}
\includegraphics[angle=-90,width=0.725\textwidth]{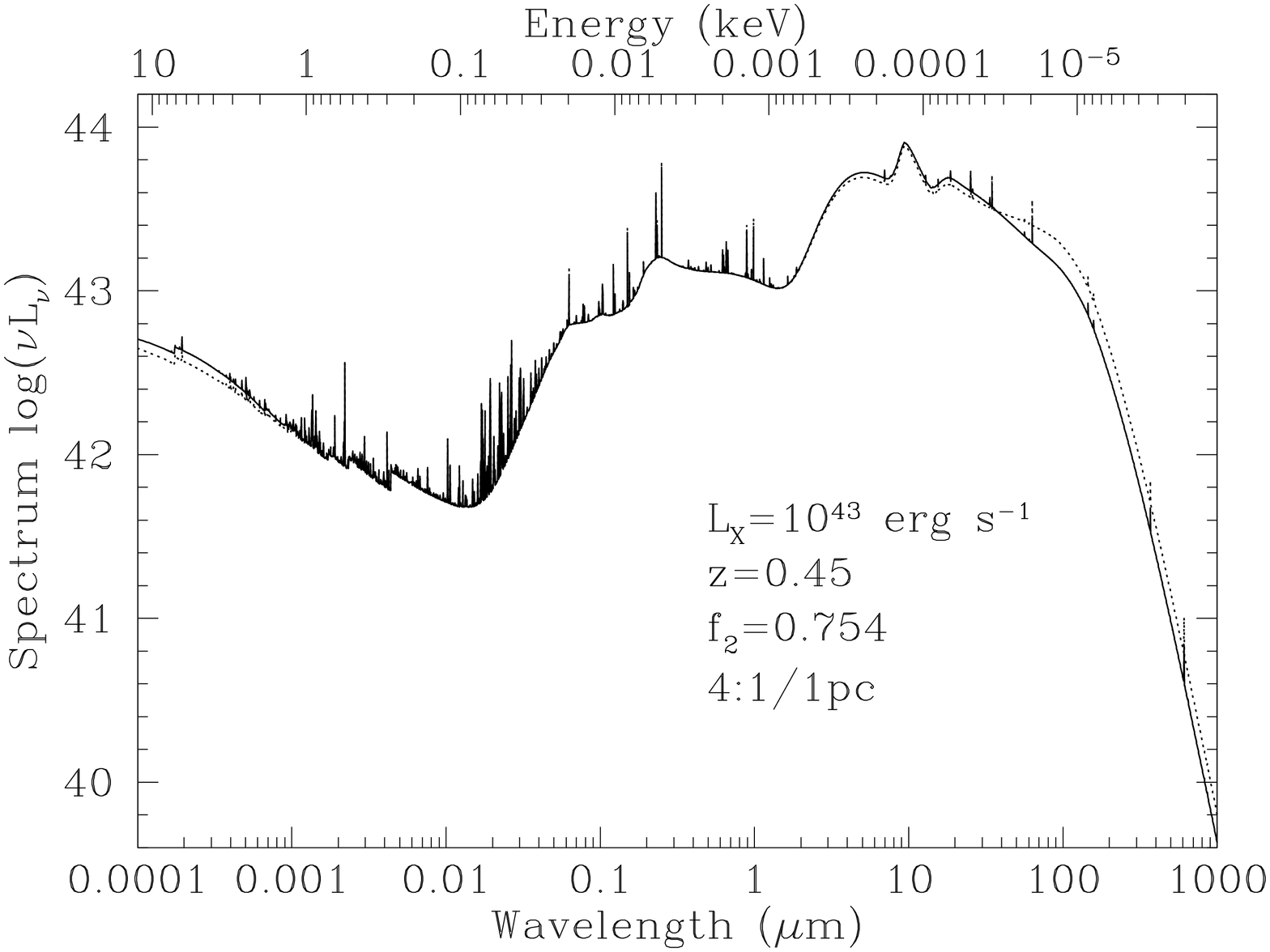}
\\
\includegraphics[angle=-90,width=0.725\textwidth]{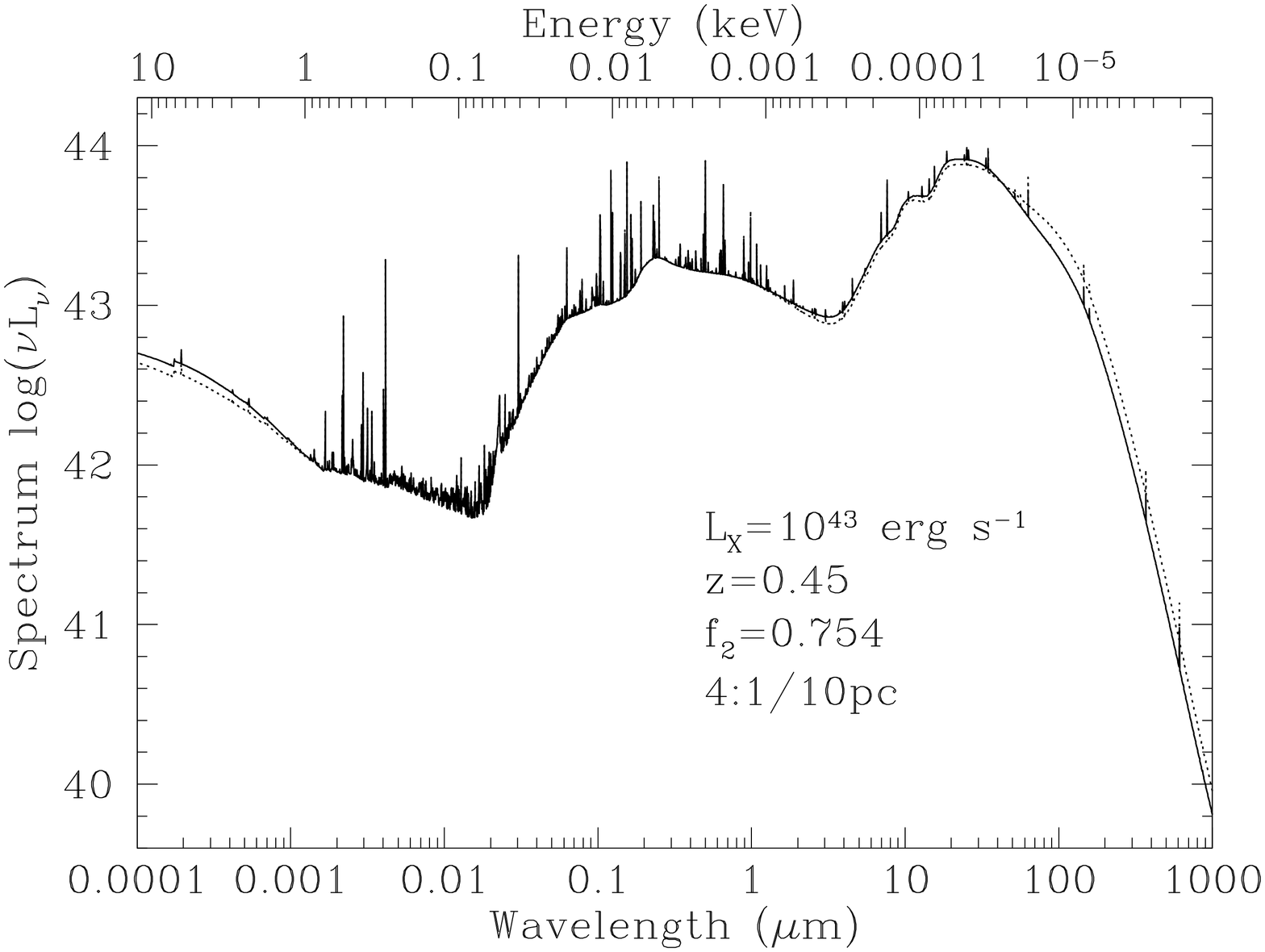}
\end{center}
\caption{(Top) Example of a \nh-averaged SED. In this
  case, the illuminating AGN has a 2--10~\kev\ luminosity of
  10$^{43}$~erg~s$^{-1}$, and the inner radius of the absorbing gas
  is 1~pc from the ionizing source. This spectrum is
  taken from the 4:1 evolutionary grid (eqn.~\ref{eq:4to1}) when
  $z=0.45$. Thus the fraction of type 2 AGN is $f_2=0.754$. The solid black line is the final
  \nh-averaged SED
  when the simple \nh\ distribution is assumed, while the dotted black
  line is the result when the Risaliti \etal\ \nh\ distribution is
  used. The Risaliti \etal\ distribution increases the emission
  longward of $\sim$30\micron, and diminishes it between $\sim 1$ and
  30\micron, as compared to the result with the uniform \nh\
  distribution. (Bottom) As above, but with $r=10$~pc.}
\label{fig:sedexample}
\end{figure}

\begin{figure}
\epsscale{1.0}
\plotone{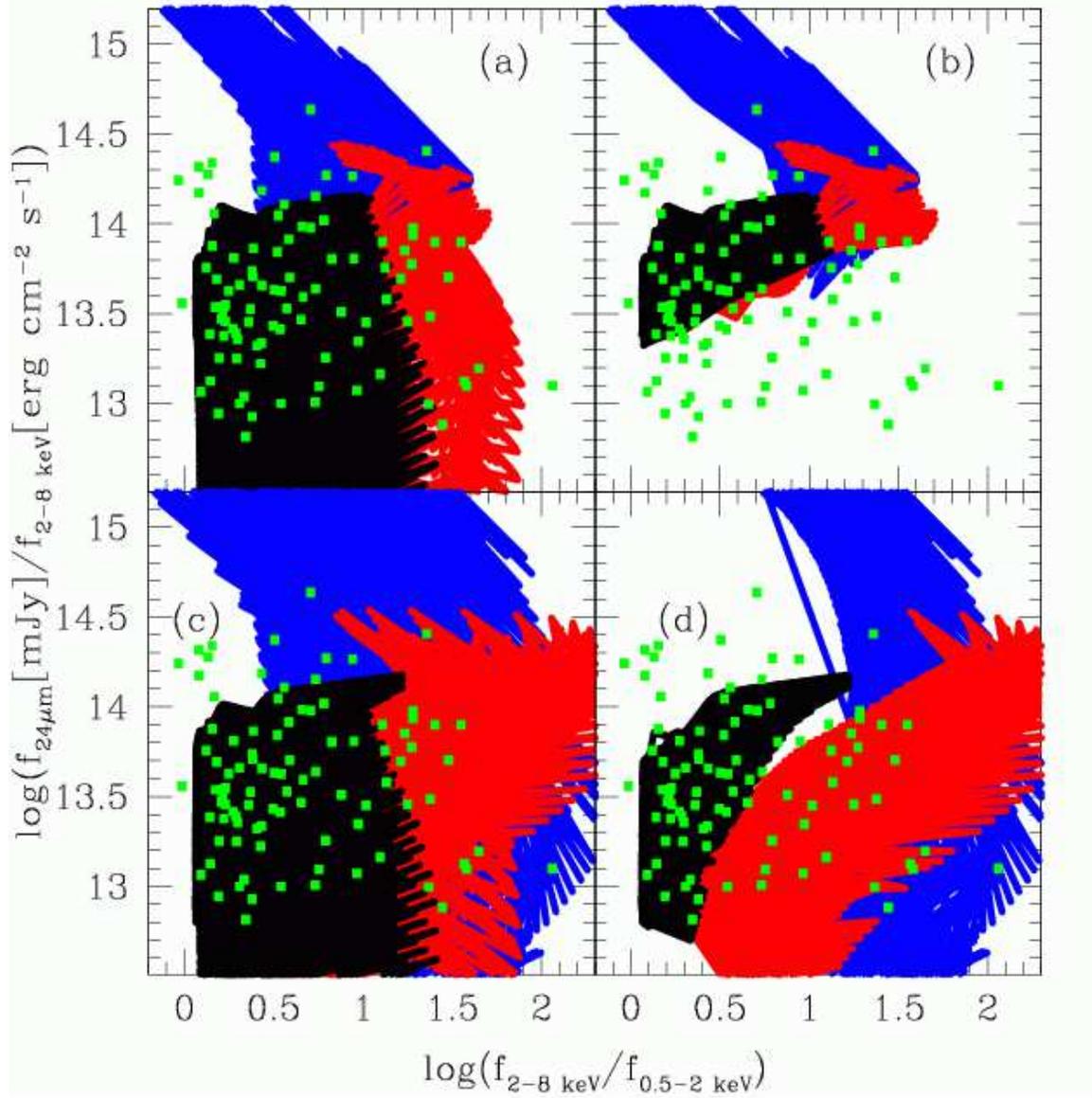}
\caption{(a) 24\micron\ to 2--8~\kev\ flux ratio versus X-ray hardness
  for all 4:1/1pc models with $z < 3$ and $\log L_{X} < 48$. The
  fluxes are taken from the unified SEDs, i.e., before the final
  average over the \nh\ distribution. The blue region covers models
  with $\log N_{\mathrm{H}} \geq 24$, the red area shows the results
  when $23 \leq \log N_{\mathrm{H}} < 24$, and the black region is derived from
  models with $\log N_{\mathrm{H}} < 23$. The green squares are data from
  the \chandra\ Deep Field South taken from Fig.\ 2 in the paper
  by \citet{rig04}. (b) As in (a) except only models with $\log L_X <
  43$ are plotted. (c) As in (a) except the results are taken from the
  4:1/10pc grid. (d) As in (b) except the results are taken from
  the 4:1/10pc series of models.}
\label{fig:nhhard}
\end{figure}

\begin{figure}
\epsscale{1.0}
\plotone{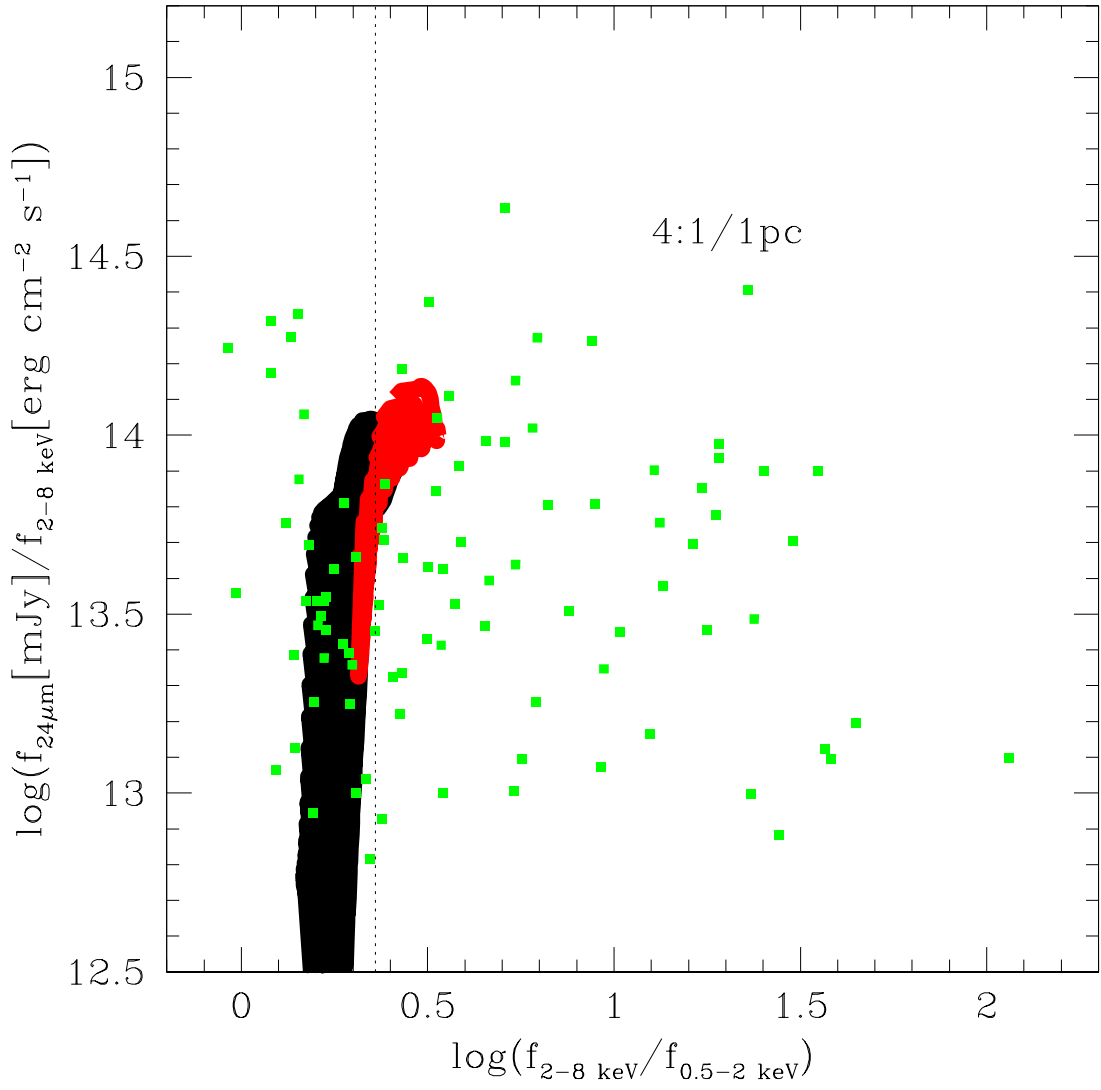}
\caption{As in Figure~\ref{fig:nhhard}(a) except the results for the
  final \nh-averaged SEDs are covered by the black region. As a result of
  the averaging, the final SEDs no longer span a large range of X-ray
  hardness. The red area corresponds to AGN with $z < 1$ and
  $\log L_X < 44$. These sources produce the largest contribution to
  the CXRB and so must have a hardness ratio corresponding to a
  $\Gamma=1.4$ power-law (i.e., $\log(f_{\mathrm{2-8\ keV}}/f_{\mathrm{0.5-2\
  keV}})=0.36$, plotted as a dotted line). Sources at
  higher redshift will appear softer.}
\label{fig:hrdratioavg}
\end{figure}

\begin{figure}
\epsscale{1.0}
\plotone{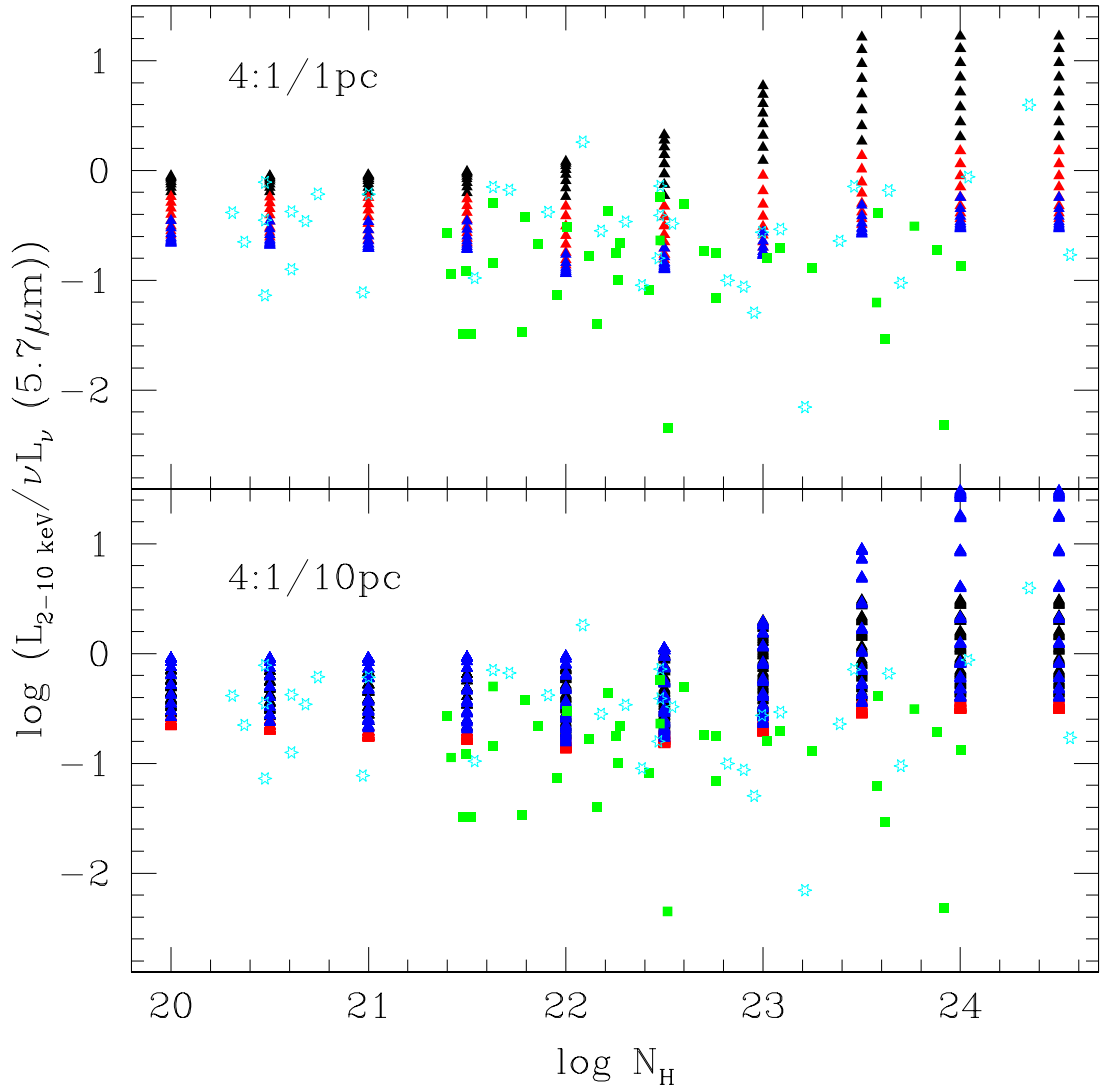}
\caption{The ratio of input $L_{X}$ to $\nu L_{\nu}$(5.7\micron) plotted
  versus $\log N_{\mathrm{H}}$ using the unified SEDs from the 4:1/1pc
  (top) and 4:1/10pc (bottom) grids. The points are color-coded by
  $L_X$: black triangles denote $46 < \log L_{X} \leq 48$, red
  triangles denote $44 < \log L_{X} \leq 46$, and blue triangles show
  results for $41.5 \leq \log L_{X} \leq 44$. Only models at $z=0.7$
  are plotted, but as rest-frame luminosities are being plotted, there
  is little change at other redshifts. The green squares
  are data from both type 1 and 2 AGN taken from \citet{rig06}, while
  the cyan stars are data from \citet{lutz04}. All of the observed
  data use 6\micron\ luminosities.}
\label{fig:rigby06}
\end{figure}

\begin{figure}
\epsscale{1.0}
\plottwo{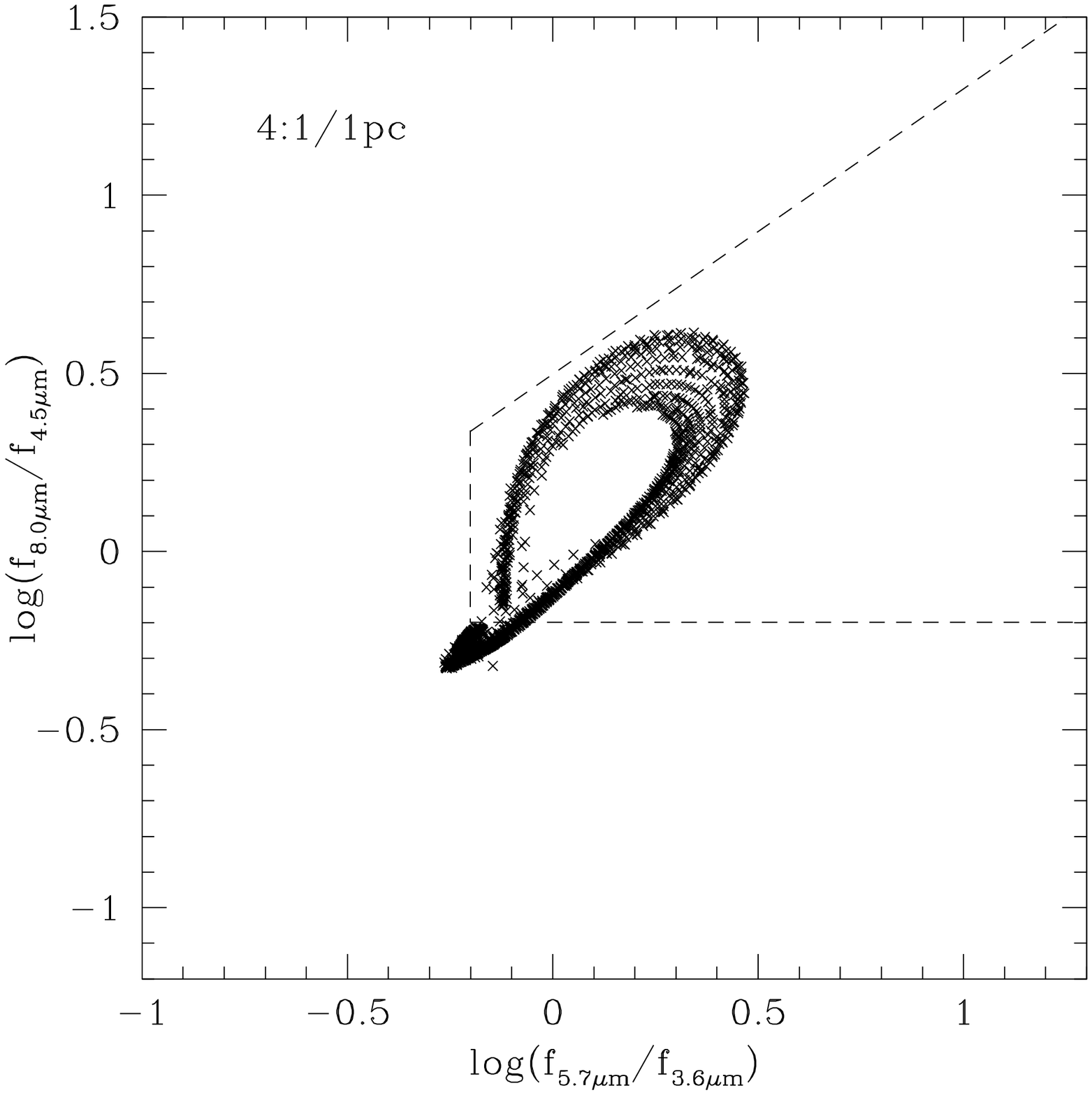}{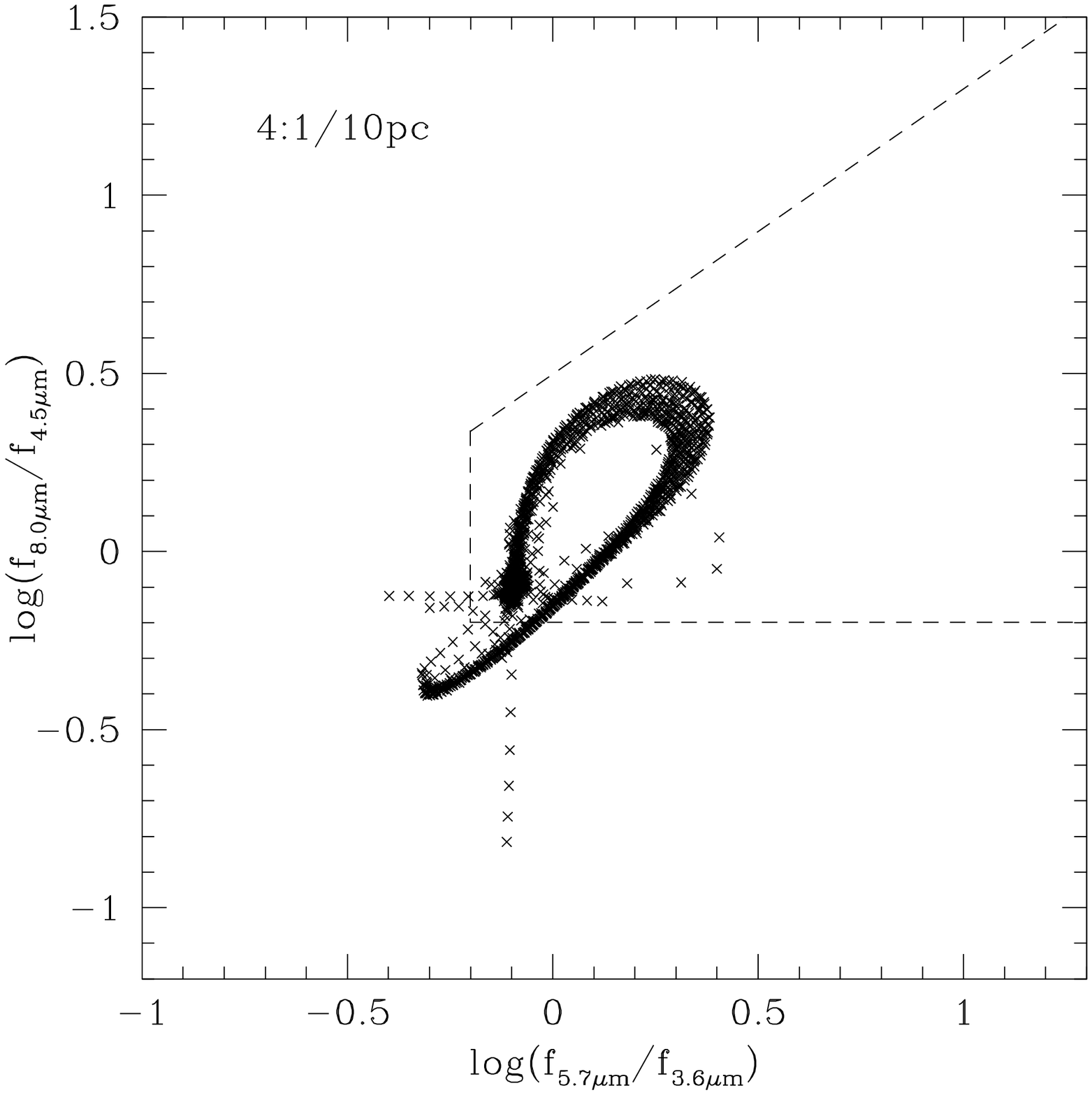}
\caption{A color-color diagram for the 4 IRAC bands from \spitzer,
  after \citet{lacy04}. (Left) The black crosses are the colors
  predicted from the 4:1/1pc \nh-averaged SEDs when $z < 3$. All
  $L_X$ are included in this diagram. The region enclosed by the three
  dashed lines is the AGN selection region used by
  \citet{lacy04}. (Right) Same as left-hand panel, but results are
  plotted for the 4:1/10pc grid.}
\label{fig:lacycolor}
\end{figure}

\begin{figure}
\epsscale{1.0}
\plottwo{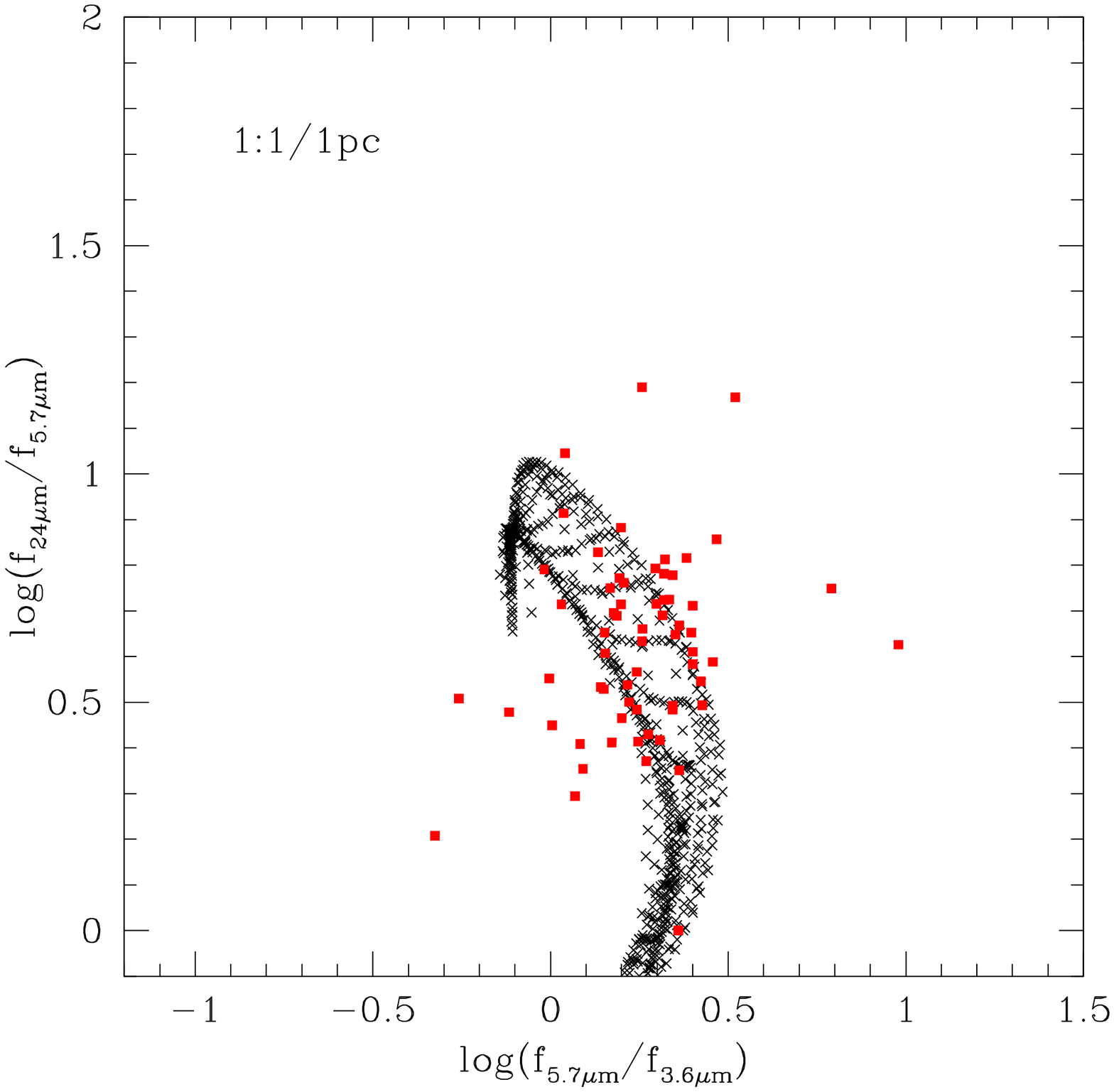}{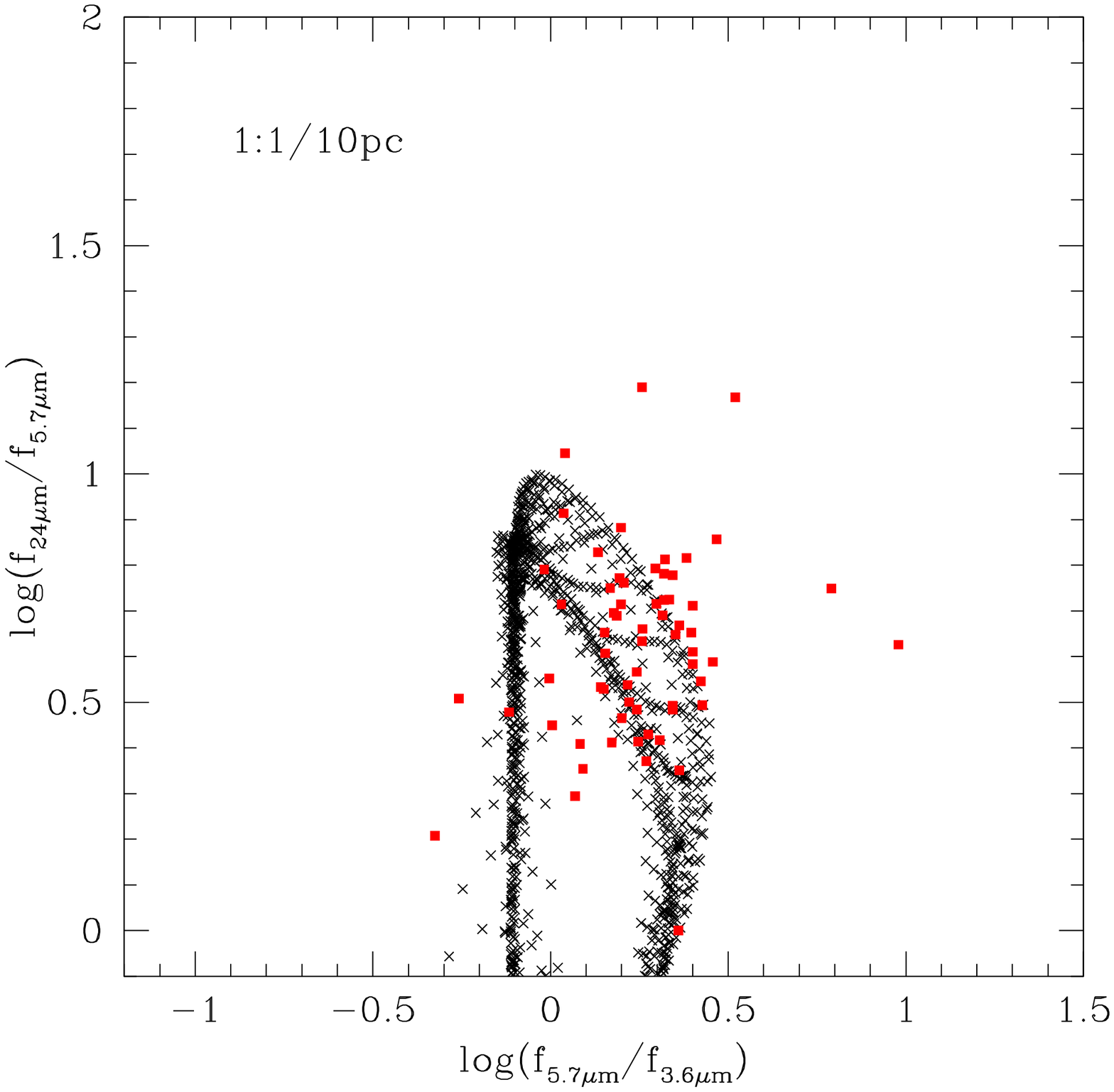}
\caption{A color-color diagram across the MIPS-IRAC bands from
  \spitzer, after \citet{lacy04}. (Left) The black crosses are the
  colors predicted from the 1:1/1pc \nh-averaged SEDs when $z <
  3$. All values of $L_X$ are included in the figure. The red squares
  are the observed colors of a variety of AGN plotted by
  \citet{lacy04} in his Figure~2. (Right) Same as left-hand panel, but
  results are plotted for 1:1/10pc grid.}
\label{fig:lacycolor2}
\end{figure}

\begin{figure}
\epsscale{1.0}
\plotone{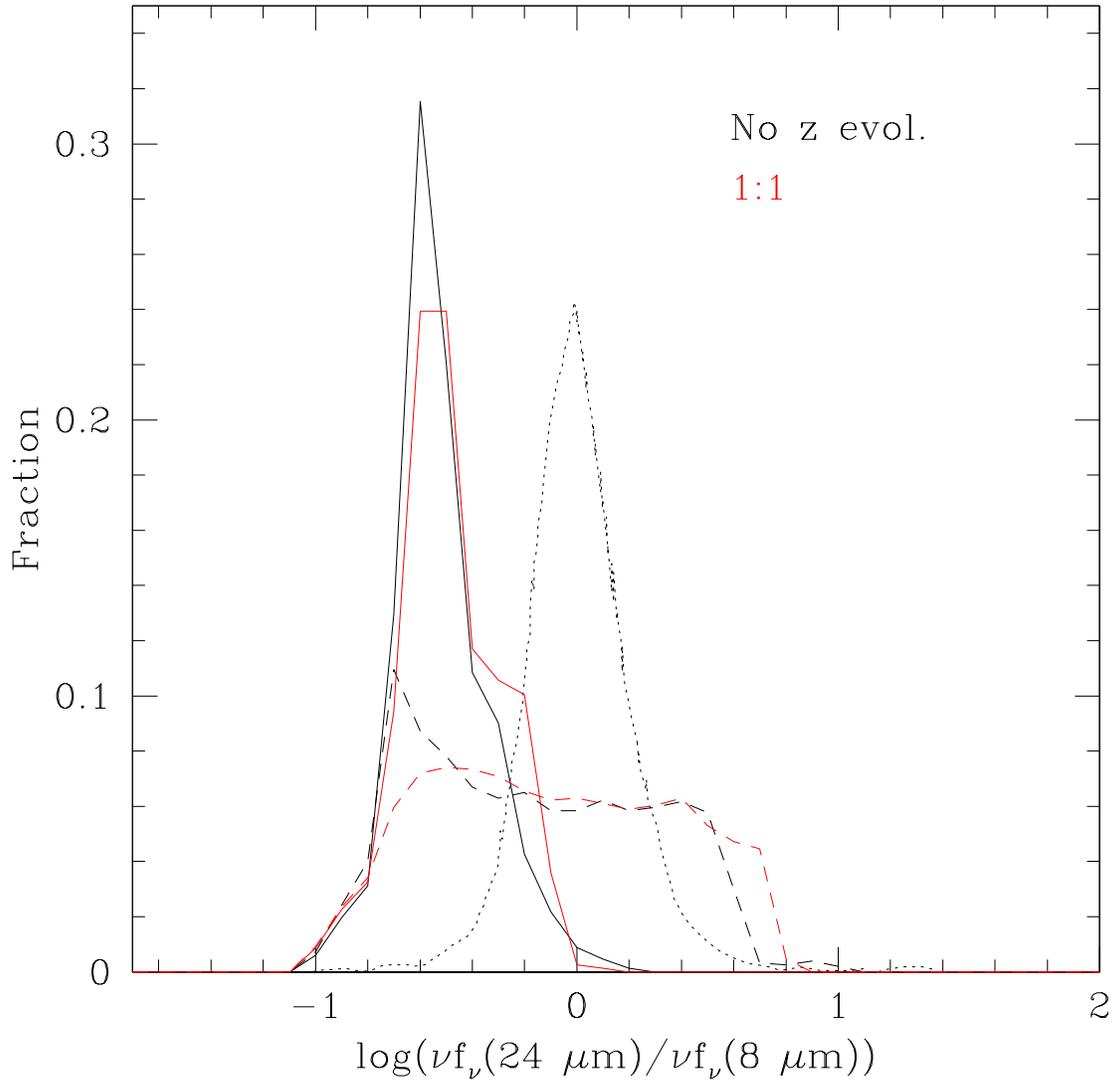}
\caption{Histogram of the 24\micron/8\micron\ flux ratio for sources in
  the nozevol/1pc (black solid line), the nozevol/10pc (black dashed
  line), the 1:1/1pc (red solid line) and the 1:1/10pc (red dashed
  line) grids. The dotted-line plots the histogram of flux ratios observed from
  X-ray selected AGN in the XBo\"{o}tes survey \citep{brand06}. Only
  the models with $f_{\mathrm{0.5-7\ keV}} > 7.8\times
  10^{-15}$~erg~cm$^{-2}$~s$^{-1}$, the flux limit of the XBo\"{o}tes
  survey, and $f_{\nu}(24\micron) > 0.3$~mJy are plotted in the figure.}
\label{fig:brandcolor}
\end{figure}

\begin{figure}
\epsscale{1.0}
\includegraphics[angle=-90,width=0.95\textwidth]{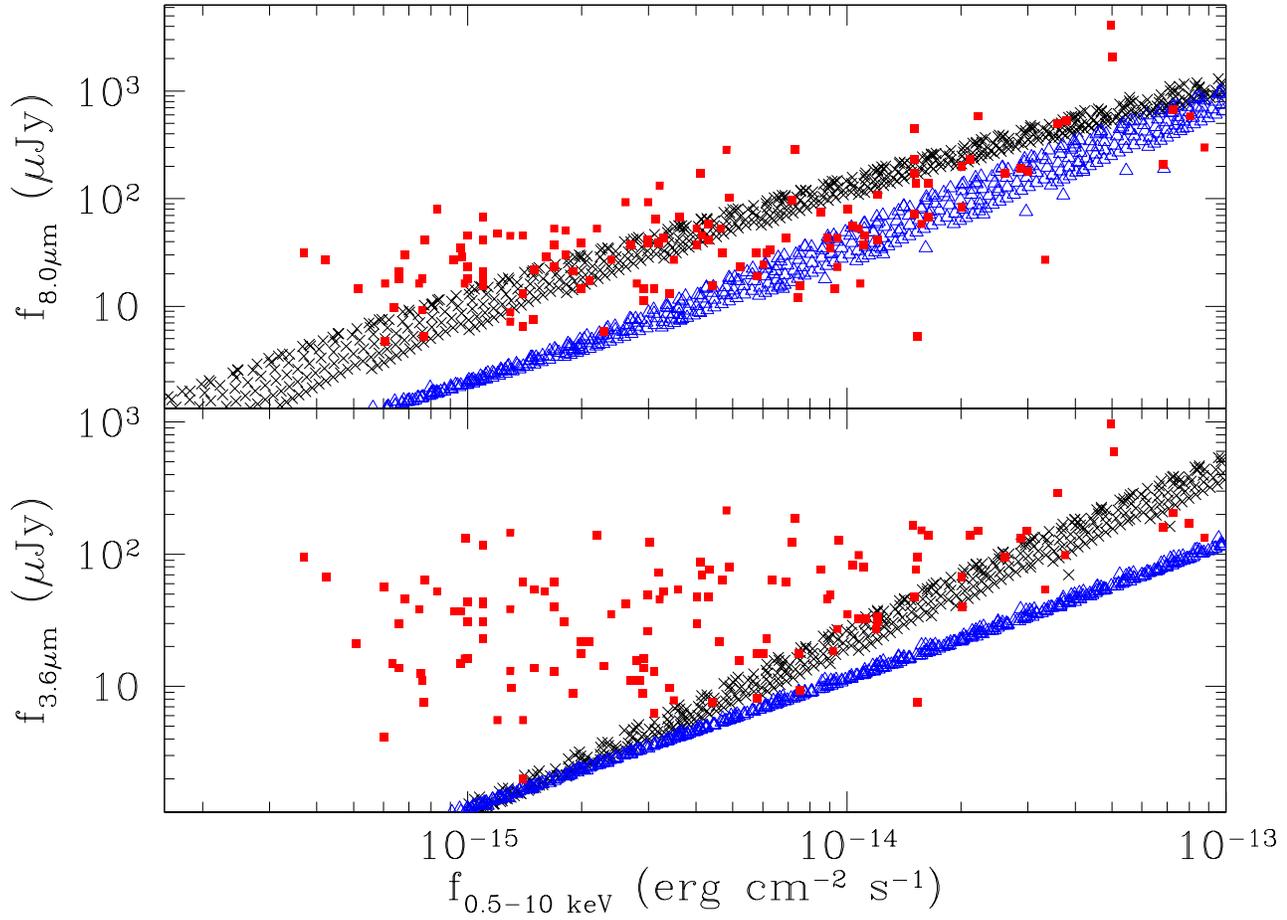}
\caption{(Top) The predicted 8\micron\ flux is plotted against the
  0.5--10~\kev\ X-ray flux. The black crosses denote results from the
  4:1/1pc grid, while the blue triangles are from the 4:1/10pc
  grid. Predictions were only plotted for $z < 3$. The red squares are
  the observed fluxes from X-ray sources in the Extended Groth Strip
  taken from Fig. 7 in the paper by \citet{barm06}. (Bottom) As above,
  except the 3.6\micron\ flux is plotted against the
  0.5--10~\kev\ flux.}
\label{fig:barmbyplot}
\end{figure}

\begin{figure}
\epsscale{1.0}
\plotone{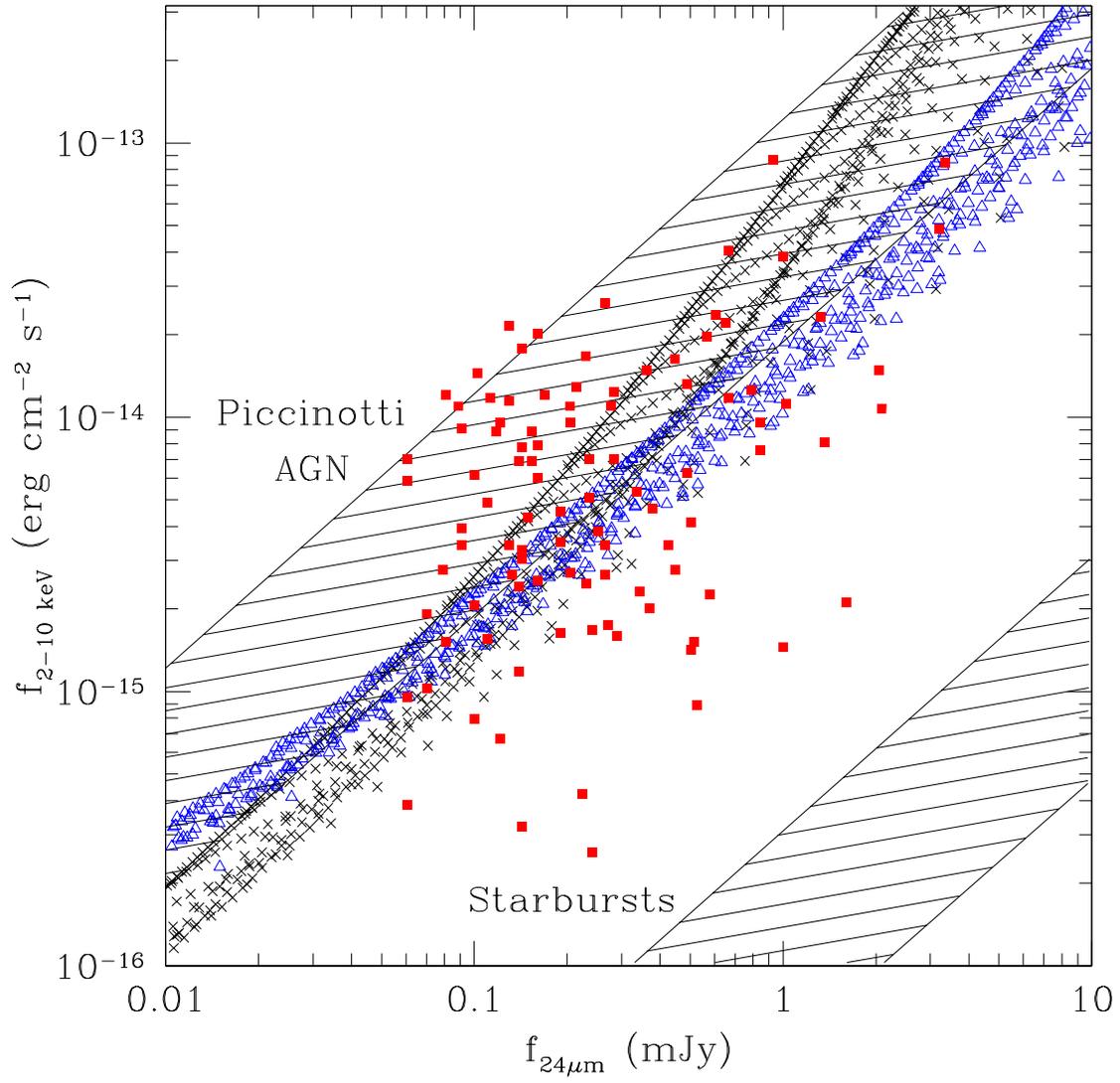}
\caption{The predicted 2--10~\kev\ flux is plotted against the
  24\micron\ flux. The symbols are
  the same as Fig.~\ref{fig:barmbyplot}, except the red squares are data
  from the \chandra\ Deep Field South taken from Fig. 1 in the paper by
  \citet{ah06}. The hatched regions denote the extrapolation of the
  hard X-ray to mid-IR ratios for low-redshift X-ray selected AGN
  (`Piccionotti AGN') and local starburst galaxies (see
  \citealt{ah04,ah06}).}
\label{fig:f2to10vsf24}
\end{figure}

\begin{figure}
\epsscale{1.0}
\plotone{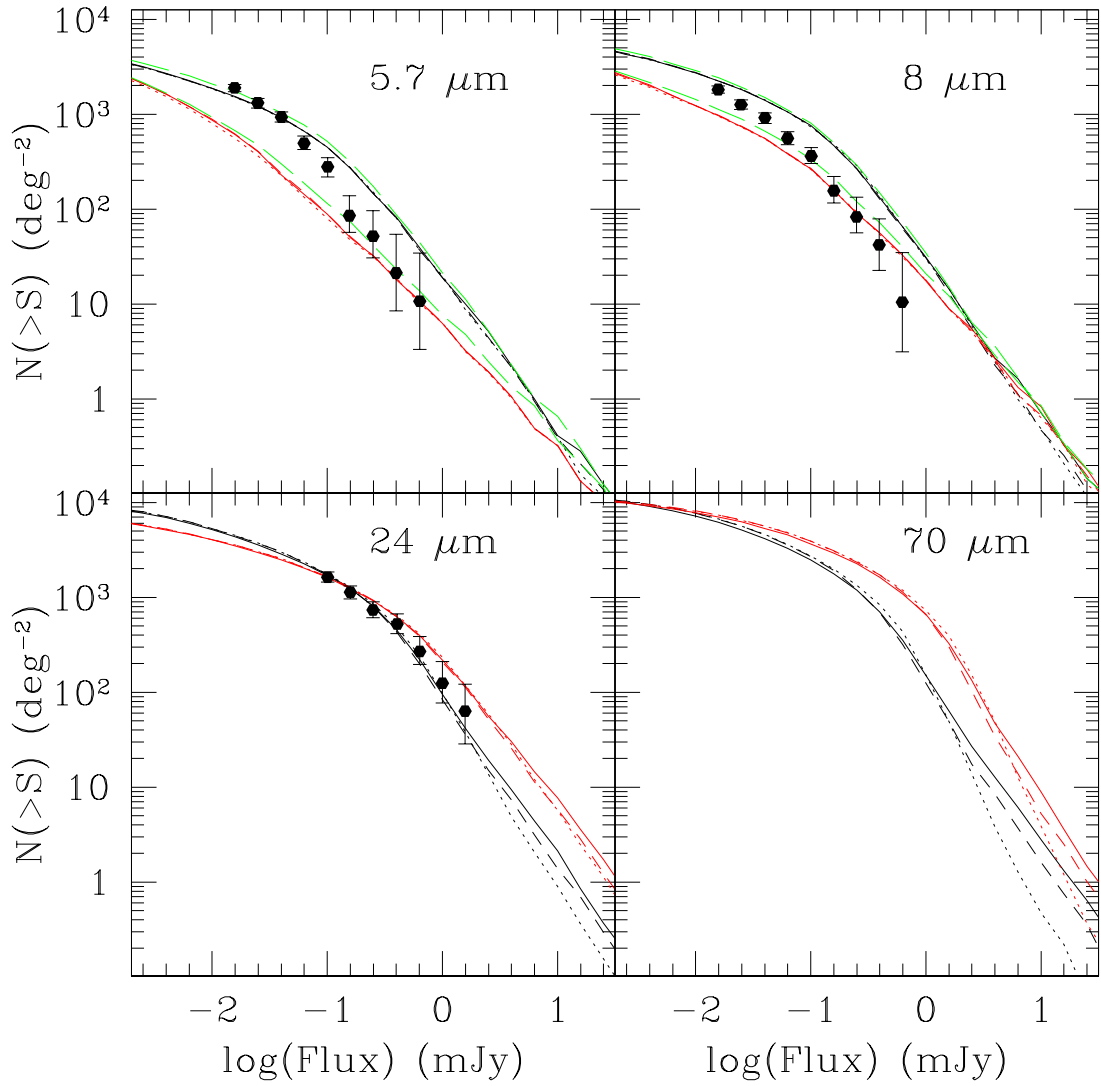}
\caption{Cumulative number count distributions for 5.7, 8, 24 and
  70\micron. The black lines plot the predicted distributions from the
  models with the inner radius of the attenuating gas and dust at $r=1$~pc from the
  ionizing source, while the red lines show the
  results when $r=10$~pc. The solid lines
  denote the no $z$ evolution model, while the dotted and dashed lines
  plot the 1:1 and 4:1 evolving models respectively. Green dashed
  lines show results for the non-evolving model computed with ISM
  grains. They are only relevant for the 5.7 and 8\micron\ panels. The simple \nh\ distribution has been used in constructing the
  distributions. The data points in the 5.7, 8, and 24\micron\ panels
  are taken from the GOODS survey of X-ray selected AGN
  \citep{tre06}.}
\label{fig:counts}
\end{figure}

As seen in Fig.~\ref{fig:contours}, the three evolutions of $R$
considered here predict very different values of the type 2/type 1
ratio with redshift. It is therefore expected that the redshift
evolution of the IR luminosity functions should provide a means to
discriminate among different evolutions of the AGN covering factor.

\begin{figure}
\epsscale{1.0}
\plottwo{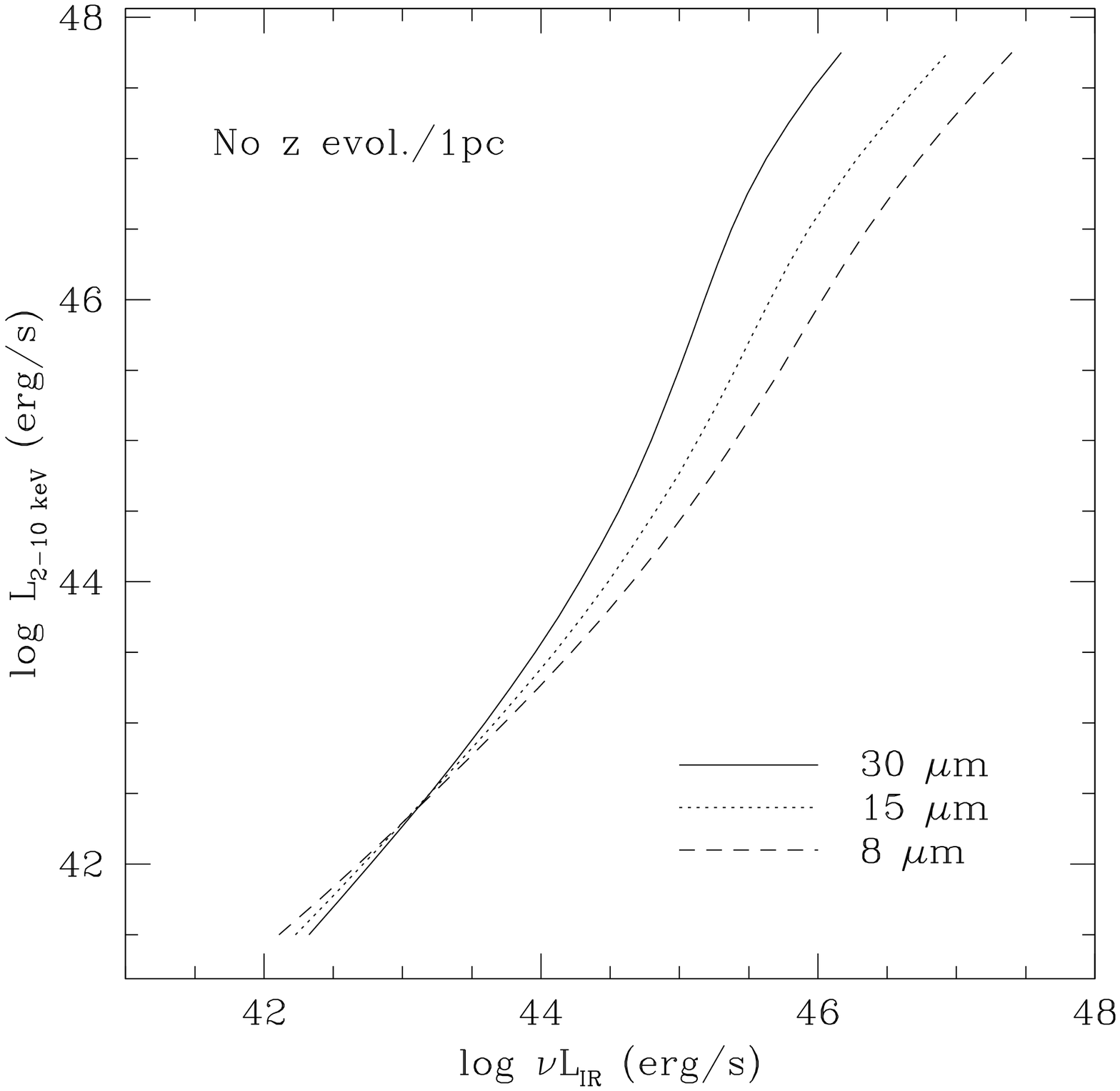}{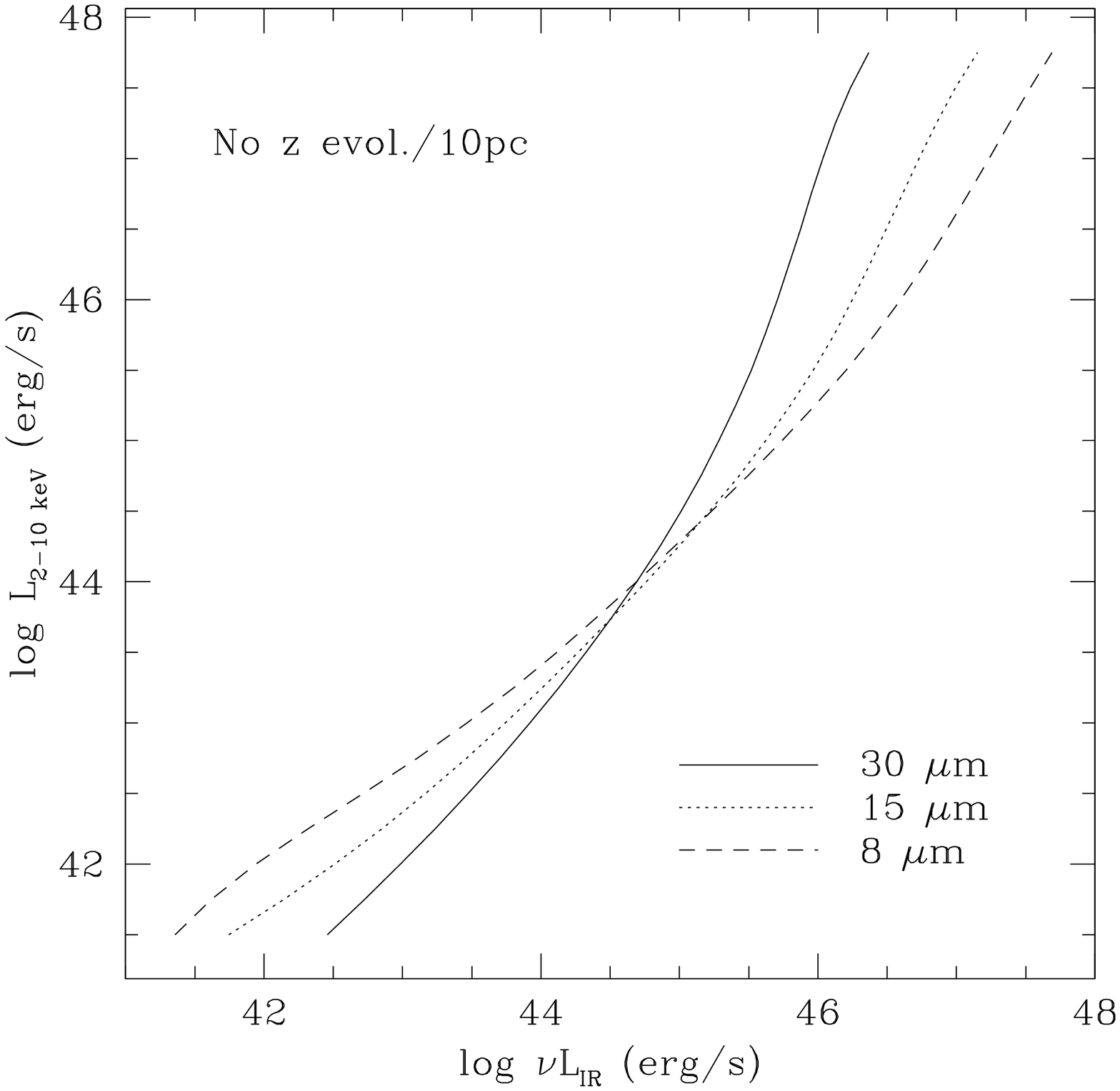}
\caption{Input hard X-ray luminosity versus predicted IR
  luminosity for the \nh-averaged SEDs in the nozevol/1pc (left panel)
  and nozevol/10pc (right panel) grids. As
  indicated in the figures, the
  various line styles denote different IR wavelengths. The range of
  luminosity spanned in the infrared is smaller than the range in the
  X-ray, with the longer wavelengths covering smaller ranges.}
\label{fig:lxvslir}
\end{figure}

The \citet{ueda03} HXLF tells us how the number density of AGN per
increment of $\log L_X$ changes with $z$ and $L_X$. The following
expression is then used to convert this to an IR rest-frame LF
$d\Phi/d(\log \nu L_{\nu})$ at a given wavelength,
\begin{equation}
{d\Phi \over d(\log \nu L_{\nu}) } = {d\Phi \over d(\log L_X)} {d(\log
  L_X) \over d(\log \nu L_{\nu}) }.
\label{eq:lf}
\end{equation}
As our photoionization technique allows a straightforward comparison
between the hard X-ray and IR luminosities for each SED, the
conversion factor in Eqn.~\ref{eq:lf} can be easily computed. As an
example, Figure~\ref{fig:lxvslir} plots the input 2--10~\kev\ luminosity
versus the IR luminosity for three different wavelengths using the
results of the nozevol/1pc and nozevol/10pc grids. The range in IR luminosities is
smaller than the range in the X-ray band, with the IR range decreasing
with increasing wavelength. For example, at $r=1$~pc, although the X-rays
span over 6 decades in luminosity, the resulting IR luminosity spans
only about 5 decades at 15\micron\ and 4 decades at 30\micron. 
This wavelength-dependent sensitivity on the input X-ray luminosity is
a result of our specific source geometry and the resulting amounts of
dust available to emit at the different IR energies. Emission at
shorter IR wavelengths is dominated by hot dust, and as the luminosity
of the ionizing source increases, greater amounts of dust are raised
to high temperature. The longer wavelengths are dominated by cooler
dust, but as the luminosity increases our models may truncate the
illuminated cloud before large amounts of this dust are warmed. This behavior is confirmed in
Sect.~\ref{sub:constantcovering}, where we show that a reduction in
the density of the IR-emitting region, and an accompanying increase in
its size, can significantly increase the long wavelength output of the
model at high $L_X$.

The factor $d(\log L_X)/ d(\log \nu L_{\nu})$ used in eq.~\ref{eq:lf}
is of course the slope of the lines plotted in
Fig.~\ref{fig:lxvslir}. At a given $z$ and wavelength, the IR LFs are
then computed for each evolutionary model by iterating over $L_{X}$,
finding the appropriate IR luminosity, determining the slope $d(\log
L_X)/ d(\log \nu L_{\nu})$ at that luminosity, and then multiplying
the HXLF at ($L_X,z$) by that slope. As with the number count
distributions, the results described below assume the simple \nh\
distribution as described in Sect.~\ref{sect:cxrbreview}.

\begin{figure}
\epsscale{1.0}
\plottwo{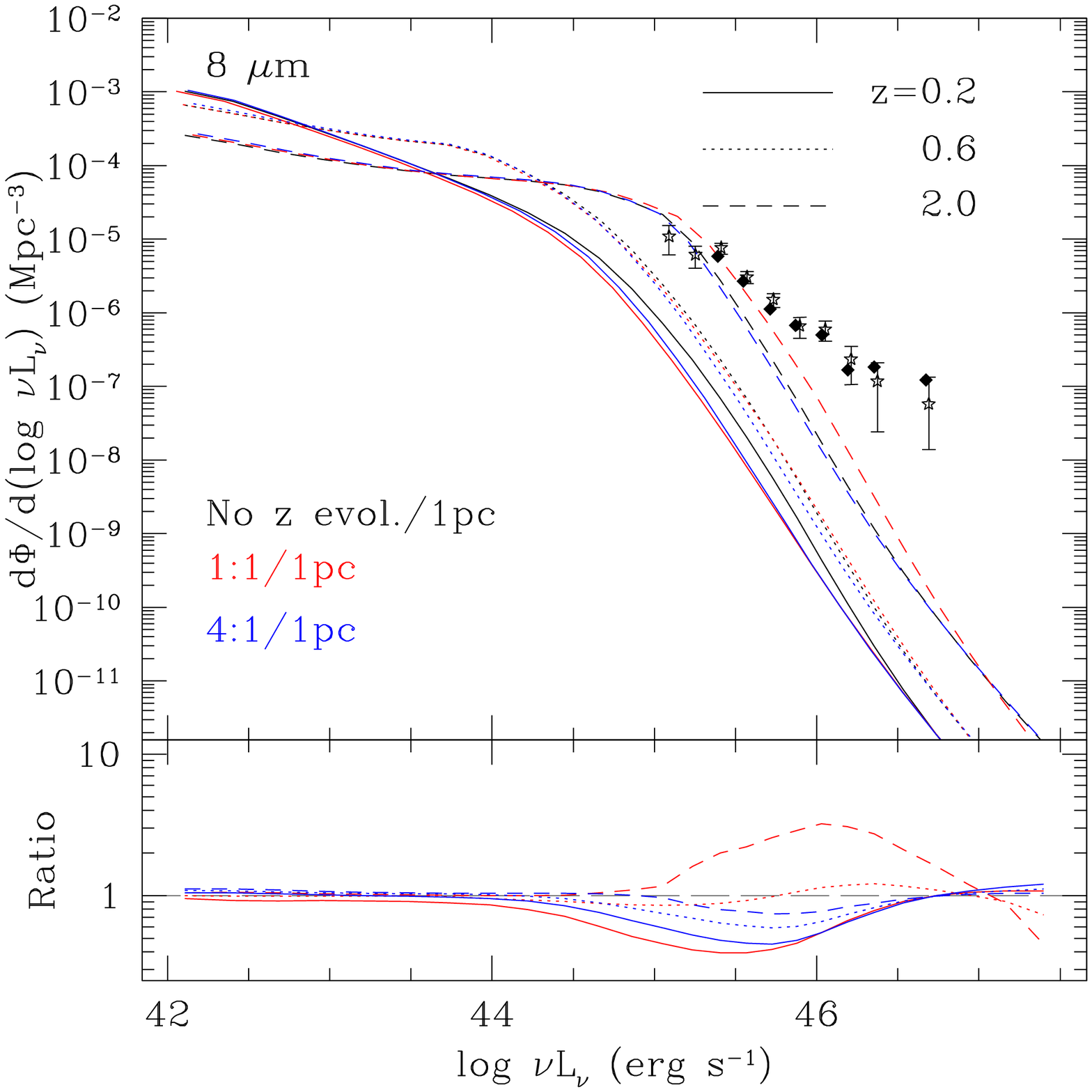}{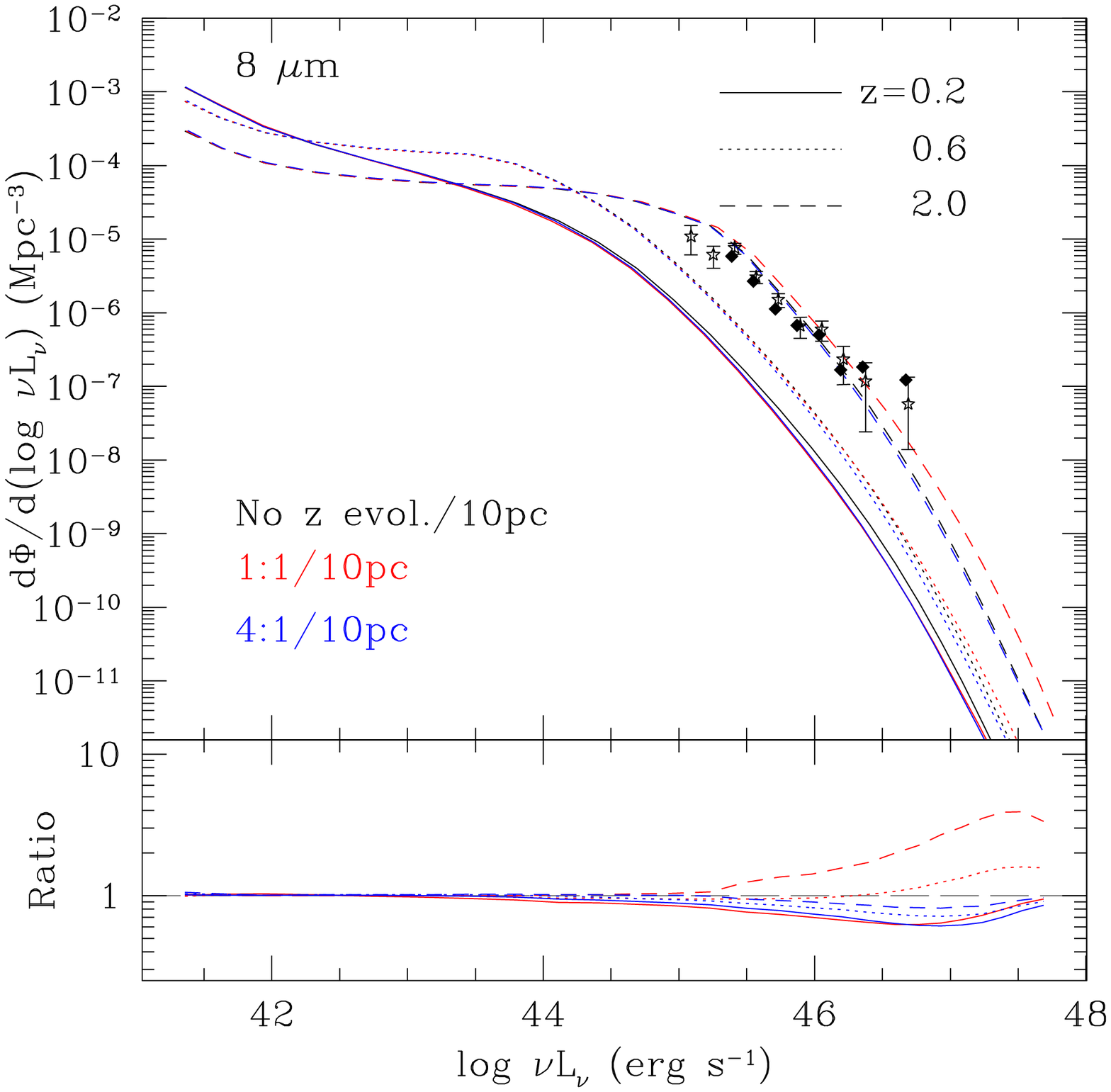}
\caption{(Left) Predicted rest-frame AGN 8\micron\ luminosity functions for the
  three different evolutions of the type 2/type 1 ratio. The black
  lines show the results for the nozevol/1pc grid, the red lines
  denote the 1:1/1pc grid, and the blue lines plot the 4:1/1pc
  results. Three different redshifts are plotted using different line
  styles. The solid, dotted and dashed lines show the luminosity
  functions at $z=0.2,0.6$ and $2.0$, respectively. The lower panel
  plots the ratio of the evolving luminosity functions to the
  non-evolving one for the three different redshifts. The colors
  denote the different evolving models as in the upper panel. For
  example, in the lower-panel the red-dashed line plots the ratio of
  the predicted 1:1/1pc luminosity function at $z=2$ to the predicted
  nozevol/1pc luminosity function at $z=2$. The data points are
  the measured luminosity function of type 1 quasars determined by
  \citet{brown06}, with the stars denoting objects with $1 \leq z
  \leq 5$ and the diamonds are objects with $1.5 \leq z \leq
  2.5$. (Right) As in the left-hand side, but for models with the
  inner radius of the 
  absorbing gas and dust at a distance of 10~pc from the AGN.}
\label{fig:8umlf}
\end{figure}

The predicted rest-frame 8\micron\ LFs for the three $R$ evolutions
are plotted\footnote{These LFs and the others at longer wavelengths
were computed using the results from grids with Orion grains. LFs
computed with ISM grains made a negligible difference, even at
8\micron, and therefore are not presented.} in Figure~\ref{fig:8umlf}
at $z=0.2,0.6$ and $2.0$. The left- and right-hand panels show the
$r=1$ and $10$~pc results, respectively. To emphasize any differences
among the LFs, the lower window in each panel shows, for each of the
three redshifts, the ratio between the evolving LFs and non-evolving
LF. It is clear that the LFs do allow a mechanism to determine among
the different evolutionary scenarios for the AGN type 2/type 1
ratio. If the inner radius of the attenuating gas and dust is on
average $\sim 1$~pc from the AGN, then, at $z=0.2$, both the 1:1 and
4:1 evolving models predict 8\micron\ LFs smaller than the
non-evolving case by factors greater than 2 for objects just past the
knee of the LF. At higher $z$, the large fraction of type 2 objects
predicted by the 1:1 evolutionary model (see
Fig.~\ref{fig:contours}(b)) results in a factor $>3$ enhancement of
the LF over the non-evolving model. If the obscuring material has an
inner radius of $\sim 10$~pc then these differences move to higher
luminosities and, at low $z$, reduce in amplitude. This is because
when the gas is further away larger AGN luminosities are needed to
produce the hot dust emission at 8\micron.

The data points in Fig.~\ref{fig:8umlf} are the measured type 1 AGN LF
at $z=2$ from \citet{brown06}. These objects were observed at
24\micron\ and lie at $1 \leq z \leq 5$ and therefore correspond to
rest-frame emission at 4--12\micron. It is obvious that the $z=2$ LFs
computed with $r=10$~pc provide a better match to the observed LF than
the ones computed with $r=1$~pc, consistent with the conclusion on $r$
reached by considering the number counts (Sect.~\ref{sub:counts}). The close match in amplitude
between the predicted and measured LFs is probably fortuitous, as the
computed LF is for both type 1 and type 2 AGNs computed at one $z$,
while the measurements are for only type 1s spread over a large range
of $z$. Nevertheless, the agreement in the shape of the LF with the
$r=10$~pc models is significant.

To explore the behavior of the absorbing material at longer
wavelengths, we plot in Figure~\ref{fig:30umlf} the predicted LFs for
rest-frame 30\micron. The differences among the three evolutions are
enhanced at this longer wavelength, with both the evolving models
underpredicting the $z=0.2$ non-evolving LF by factors $>
10$. Moreover, for both the $r=1$ and $10$~pc set of models the
differences persist over a much larger range of luminosity than at
8\micron. At $z=1$, the large number of obscured AGN predicted by the
1:1 evolution leads to an enhancement of the 30\micron\ LF by factors
that can approach 50 if $r=1$~pc or 100 if $r=10$~pc. The largest
differences among the evolutions appear above the knee of the LFs and
move to higher luminosities in the $r=10$~pc models. Evidently, the
warm dust emission is very sensitive to the covering factor of gas
around the AGN, and therefore provides a clear tracer of any evolution
with $z$.

In order to compare these computed 30\micron\ LFs to data we made use
of the SWIRE survey of \citet{fra05}. These authors presented AGN SEDs
from \spitzer\ IR to optical and X-ray wavelengths for X-ray detected
AGN in the ELAIS N1 region.  Using their SEDs, we estimated the
60\micron\ flux density and computed the rest-frame 30\micron\ LF at
$z=1$.  The LF was derived by the $1/V_{\mathrm{max}}$ method. The
redshift distribution of sources peaks around $z\sim 1$ and the
derived LF is not sensitive to the choice of the redshift bin.  A
redshift bin of 0.8--1.2 was selected in order to ensure a reasonable
number of objects (the LF changes by $<20$\% for a
redshift bin 0.9--1.1). The range of 30\micron\ luminosity, $\nu
L_{\nu}$, of objects at $0.8 < z < 1.2$ is about one order of
magnitude and we employed a range of $\nu L_{\nu} =
10^{44}$--$10^{44.5}$~erg~s$^{-1}$ to exclude an outlier. For
simplicity, we assume the same incompleteness over both the redshift
and luminosity bins, although this can result in overestimating the
LF.  However, the LF without the incompleteness correction is smaller
by only 30\%, within the estimated uncertainty. Corrections were made
for the 99\% completeness of X-ray detection \citep[see][]{Manners03},
the 90\% completeness of infrared source detection, and the
identification of 102/122 X-ray sources in the \spitzer\ IRAC
image. The uncertainty of the LF includes the Poisson noise statistics
on the number of sources and the uncertainty in the observed
30\micron\ luminosity estimated by fitting the SEDs. While the IR SED
fitting is based on the IRAC data, most of the objects used for
constructing the LF also have 24\micron\ detections and they are
roughly consistent with the SED fitting. The
$f_{\mathrm{25\micron}}/f_{\mathrm{12\micron}}$ ratio of local
($z<0.2$) Seyfert galaxies varies only by a factor of $\sim$3.5
\citep{Neugebauer86}. Thus, we argue that this factor is a
conservative estimate of the uncertainty in the observed 30\micron\
luminosity, because the IR SED fitting performed by \citet{fra05} is
based on IR data points at several wavelengths.  The final LF is
plotted in Figure~\ref{fig:30umlf}, where the uncertainty is dominated
by the uncertainty in the IR luminosity.

This datapoint is consistent with the predictions of the $r=1$ and
$r=10$~pc models. If the rest-frame 30\micron\ emission from an
average AGN with this luminosity at $z=1$ arises from material with an
inner radius of $\sim 1$~pc from the central source, then only the 1:1
evolution predicts enough absorbed AGN to account for the observed
LF. Converting the IR luminosity of this data point to an X-ray
luminosity gives $\log L_{X} \approx 44$. At this luminosity and
$z=1$, Fig.~\ref{fig:contours} shows that evolution~\ref{eq:1to1}
predicts a type 2/type 1 ratio of $\sim$5:1, as opposed to the $\sim$
2:1 ratio derived from the other two models. These lower ratios are
consistent with the data point if the inner radius of the absorbing
gas and dust is at 10~pc. Unfortunately, in this situation the
luminosities of the AGN are too low to distinguish among the three
evolutions.

Moving to even longer wavelengths further enhances the differences
among the three different models of $R(L,z)$, although complications
arising from star formation are more important at longer wavelengths
(see next section). The rest-frame 35\micron\ LF, which is accessible
to \spitzer\ at $z=1$ using the 70\micron\ MIPS band exhibits a
similar behavior as seen at 30\micron\ with the differences in the
LFs increased to factors $\sim 100$ at both $z=0$ and $1$.

\section{Accounting for Star Formation}
\label{sect:sf}
In the previous section we found that large differences among
the three type 2/type 1 evolutions are found at 30\micron\ (for both
the 1~pc and 10~pc models). However, these longer wavelengths will
likely be contaminated with emission from star-forming regions. This
will be particularly troublesome at $z \sim 1$ if the AGN absorbing
gas is connected to the SFR in the host galaxy. Therefore, some limits
need to be placed on the effects of star-formation on the above
predictions.

A simple way to proceed is to make use of the work of \citet{yc02},
who fit a single temperature dust model to the far-IR/sub-mm emission
of local starburst galaxies (they also included bremsstrahlung and
synchrotron emission, but emission from these processes is only important
at much longer wavelengths). The median dust temperature
$T_{\mathrm{d}}$ and emissivity index $\beta$ they find from their
sample is $T_{\mathrm{d}} = 58$~K and $\beta = 1.35$. By relating the
total 40-500\micron\ emission to the SFR by SFR$=
L_{\mathrm{40-500\micron}}/(5.8\times 10^{9}
\mathrm{L_{\odot}})$~M$_{\odot}$~yr$^{-1}$, \citet{yc02} then write
the dust emission as
\begin{equation}
L_{\nu} = 1.56\times 10^{21} \nu^3 \mathrm{SFR}
{(1-e^{-(\nu/2000)^{1.35}}) \over (e^{0.00083\nu}-1)}\ \ \mathrm{erg\ 
  s^{-1}\  Hz^{-1}},
\label{eq:sf}
\end{equation}
where $\nu$ is in GHz.

As this spectrum assumes only one relatively cool dust temperature and
ignores more complicated dust physics such as PAH emission and
non-linear temperature spiking from very-small grains, it will not be
an accurate correction to the AGN spectra at wavelengths $\lesssim
20$\micron. It should be a reasonable correction at 30\micron.  The dust
emission spectrum (Eq.~\ref{eq:sf}) was added to the final \nh-averaged
\cloudy\ SEDs between 3 and 1500\micron\ assuming constant SFRs of
0.5, 1, 5, 10 and 100~M$_{\odot}$~yr$^{-1}$.

As the number count distributions did not provide a good method to
discriminate among the three evolutionary models, we will omit a
discussion on how star-formation impacts the number counts. Suffice it
to say that the only non-negligible effect on the predicted
distributions occurs at 70\micron.

Turning to the LFs, Figure~\ref{fig:30umlf-sf} plots how one particular
30\micron\ luminosity function is altered by adding in dust emission
from the different SFRs. For SFRs $\leq 10$~M$_{\odot}$~yr$^{-1}$
significant deviations occur only when $\log \nu L_{\nu} \lesssim
44.5$. At low AGN luminosities the majority of the 30\micron\ emission
is from the star-formation emission. In contrast, the 30\micron\ LF
can be dramatically changed at nearly all luminosities if the SFR
$\sim 100$~M$_{\odot}$~yr$^{-1}$. When the SFR reaches these values,
the 30\micron\ luminosity is only mildly sensitive to the heating from
the AGN. Although there is significant scatter from object to object,
such a high SFR seems to be reached in an average galaxy only at $z
\gtrsim 1.5$ \citep{jun05,pap06}. However, in spite of the changes to
the 30\micron\ LF caused by star-formation,
Figure~\ref{fig:30umlf-sf2} shows that the LFs from the three AGN type
2/type 1 ratios are still significantly different for SFRs $\leq
10$~M$_{\odot}$~yr$^{-1}$, and the discussion from the previous
section is still valid. This figure does suggest, however, that
generating a LF from sources with a lower SFR would be preferable.

\section{Discussion}
\label{sect:discuss}
The goal of this paper was to investigate how three possible
evolutions of the AGN type 2/type 1 ratio (all of which fit the
CXRB) manifest themselves in the mid-IR. The predictions were computed
by employing the photoionization code \cloudy\ to calculate the SED
from an obscured AGN over a large range of X-ray luminosity, $L_{X}$,
and redshift.  In this way, a direct connection between the observed
X-ray and IR fluxes could be predicted. The covering factor of the
absorbing gas and dust changed with $z$ and $L_{X}$ as predicted by
CXRB synthesis models and the unified model. 

The results presented in the above two sections show that this
photoionization-based method is very successful. The family of
computed SEDs indeed capture many of the average properties in typical
AGN observed in the deep surveys of \chandra, \spitzer\ and \xmm,
including infrared colors and observed IR and X-ray fluxes. Typically,
investigations of the IR properties of AGN involve either detailed
dust radiative transfer models \citep[e.g.,][]{gra97} that provide no
connection to the X-rays, or are based on phenomenological SEDs and
are unable to provide constraints on physical parameters
\citep[e.g.,][]{silva04}. Our approach is not designed to fit the SEDs
of individual objects, but we can use it to constrain physically
interesting quantities in the ensemble of objects. In this way, a
picture of an average AGN can emerge, uncluttered from the significant
source-to-source variability that affects the study of small samples.

\subsection{The Location of the Absorbing Gas}
\label{sub:radius}
Two different inner radii were used in computing the \cloudy\ models,
1~pc and 10~pc (see \S~\ref{sub:inner}), as tests for whether large
samples of AGN show evidence for dust emission at these average
distances. Although only these two radii were considered here, comparing
the results from the two sets of models with the available data gave
rise to the interesting result that almost all observational tests
preferred the 10~pc models over the 1~pc ones (Figs.~\ref{fig:nhhard}
and~~\ref{fig:brandcolor}). In particular, the GOODS number-count
distributions were best described by the 10~pc models from
observed-frame 5.7\micron\ to 24\micron\ (Fig.~\ref{fig:counts}).

It is important to note here
that both the $r=1$ and $10$~pc sets of models were assumed to have
the same constant density of $10^4$~cm$^{-3}$. Thus, all \cloudy\ runs
with $\log N_{\mathrm{H}} < 23.5$ have pathlengths less than 10~pc
(ignoring ionization effects). As
a result the $r=1$~pc models will be dominated by hot dust, as the
column does not extend to great enough distances to produce
significant warm emission. A new lower density distribution that would
have an inner radius of 1~pc and be able to produce the correct
24\micron\ flux is certainly conceivable. On the other hand, if the
obscuring medium is clumpy with each clump being roughly the same
density \citep{dv05}, then our setup may not be too unrealistic. An
investigation using different density distributions will be the
subject of future work.

Although the 10~pc model seems to be the best description of data
observed at the mid-IR \spitzer\
wavelengths, the observed SEDs of many AGN show that hot dust from
inside this radius is also required. The 10~pc
\cloudy\ models do predict dust temperatures close to and at the sublimation
value at the inner radius of the cloud, but only for the more luminous
AGN models (see \S~\ref{sub:inner}). Thus, there must be dust emission over a
range of radii and temperatures, in agreement with SED fitting of
individual objects \citep[e.g.,][]{kur03}. This conclusion is also
consistent with observations searching for variability in the IR continua
of AGN that find little to no variations at mid- to far-IR
wavelengths on day to month long timescales
\citep{rl81,em87,cle88}. In contrast, near-IR (JHK band) variability
of local Seyferts is observed with measured
timelags from the optical/UV placing the emitting region $< 1$~pc from
the continuum source \citep{cla89,sug06,min06}.

\citet{nm99} report long term monitoring of nearby, low-luminosity
quasars at 10.6\micron. At the typical redshifts of their sample, this
wavelength is very close to 8\micron\ in the rest-frame. They find
slow, multiple-year variability but conclude that in no case are the
results in contradiction to a thermal origin for the emission in the
objects with luminosities $< 10^{13}$ L$_\odot$. Specifically, the
variability time scales are consistent with our conclusion that the
8\micron\ emission arises from thermal re-radiation by dust in a
component with a typical inner radius of $\sim 10$~pc. They also
summarize previous measurements of luminous, radio-loud quasars, such
as 3C~273; in this object in particular, observations at $\sim
8$\micron\ in the rest frame show variations that are too fast to
interpret as thermal radiation. Since such objects are exceptional
among the total quasar population, it is unlikely that such behavior
would strongly affect our results for the ensemble properties of these
sources.

In the future, model grids can be extended to even larger inner radii
to establish a pair of bracketing values for this key parameter. It
will also be possible to break observed data into redshift and
luminosity slices to determine if there is an evolution of the inner
radius with $z$ and/or $L_{X}$ \citep[e.g.,][]{hatz06}.

\subsection{Determining the Evolution of $R$}
\label{sub:evol}
Sections~\ref{sect:agn}~and~\ref{sect:sf} showed that the mid-IR LF of
AGN will be able to discriminate among different models of $R$, the
type 2/type 1 ratio. Out of the wavelengths considered here, the rest
frame 30\micron\ LF between $z=0$ and $1$ gives the largest
differences among evolutionary scenarios, as well as being the least
contaminated due to star-formation emission. This last point needs to
be confirmed with a more sophisticated prediction of the IR emission
due to star-formation.

\begin{figure}
\epsscale{1.0}
\plottwo{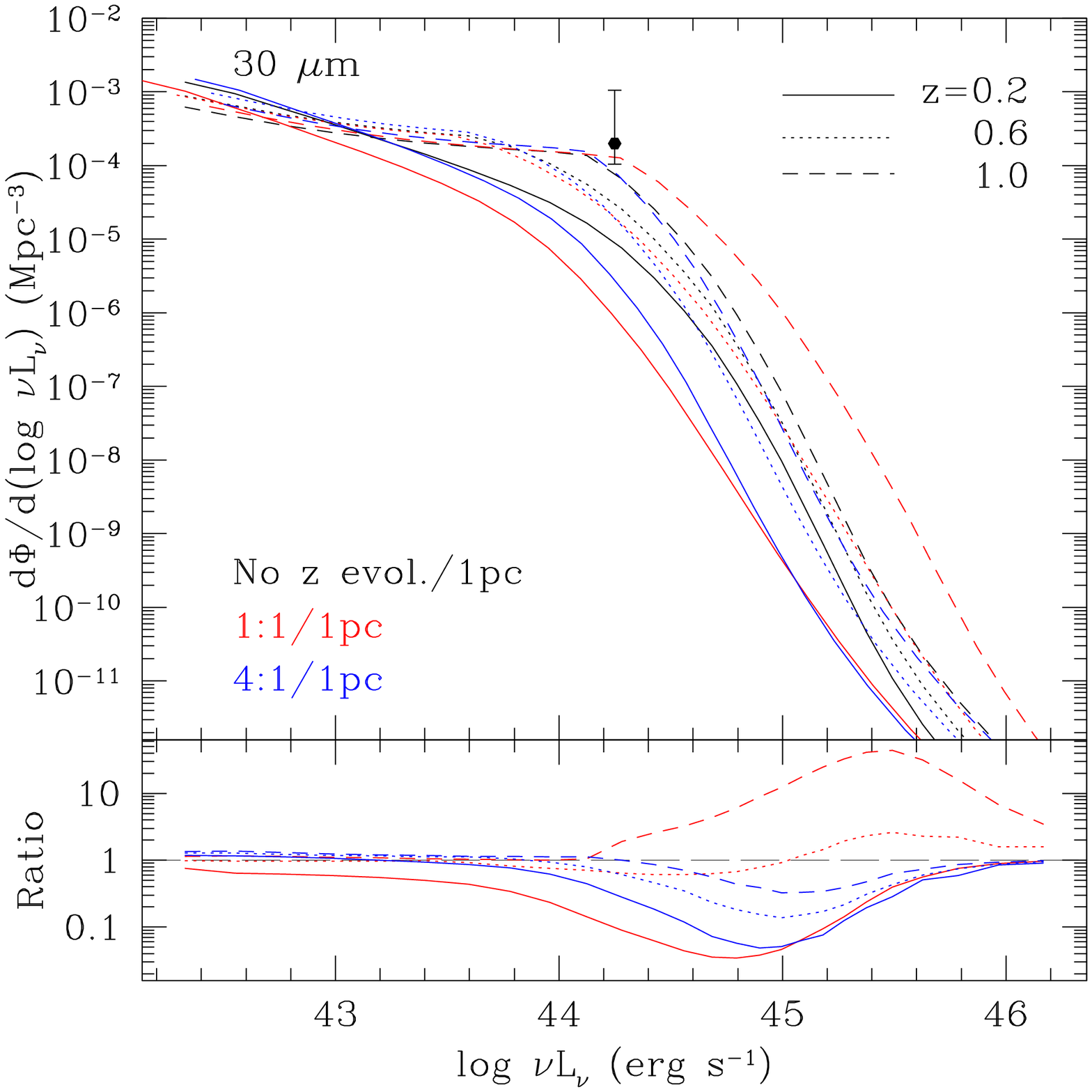}{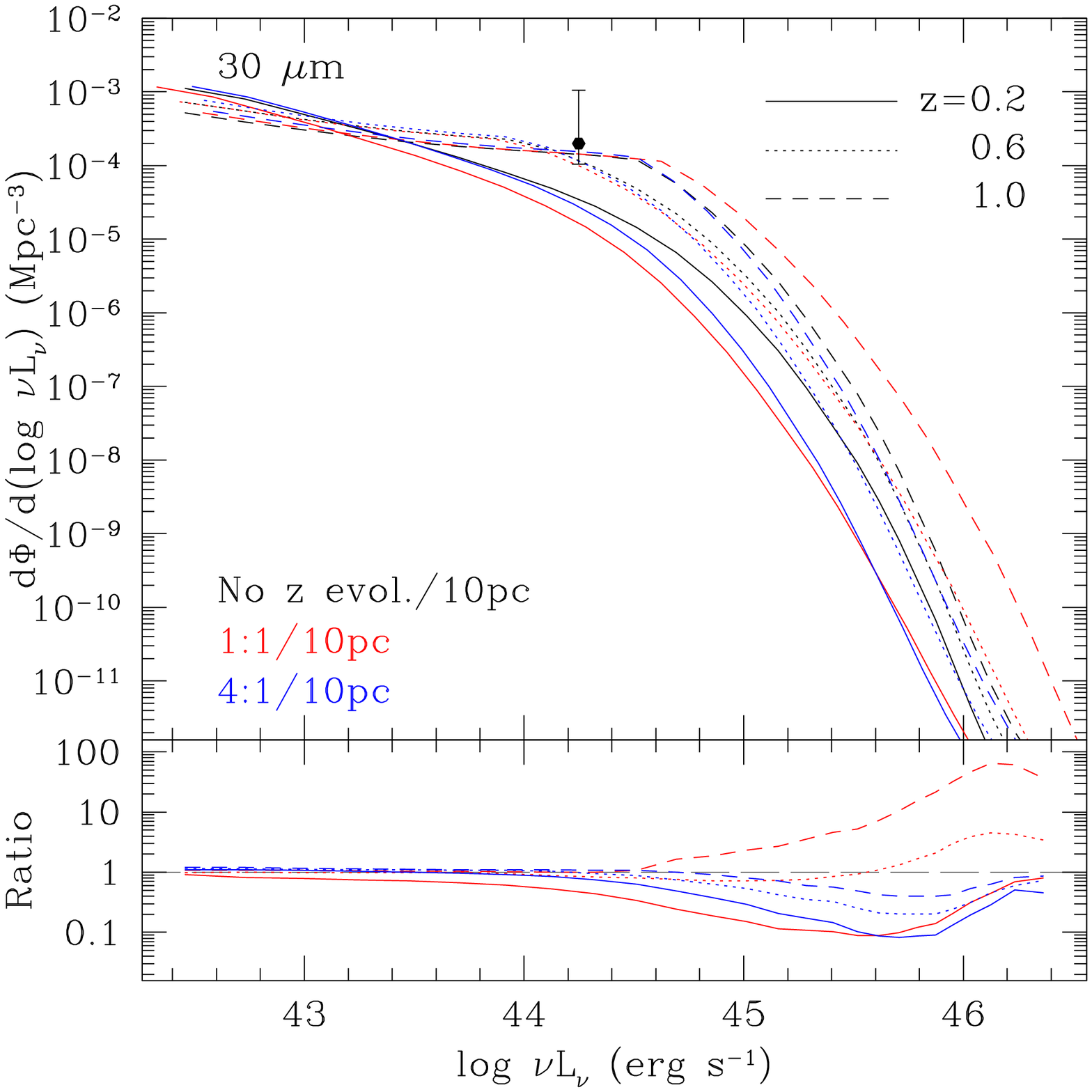}
\caption{As in Fig.~\ref{fig:8umlf} except for the rest-frame
  30\micron\ luminosity functions. The data point is calculated from the
  SWIRE survey of \citet{fra05}. See text for details.}
\label{fig:30umlf}
\end{figure}

\begin{figure}
\epsscale{1.0}
\plottwo{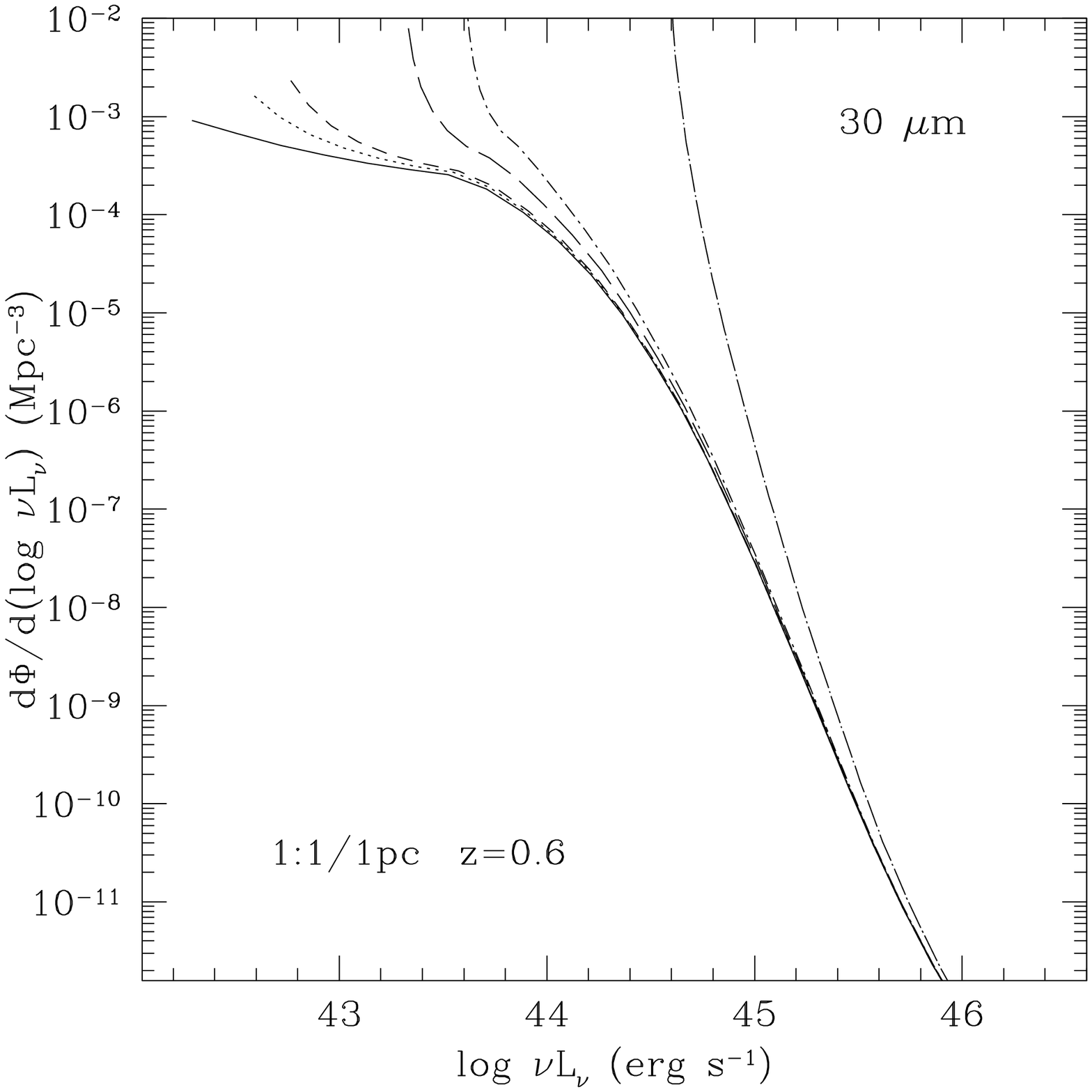}{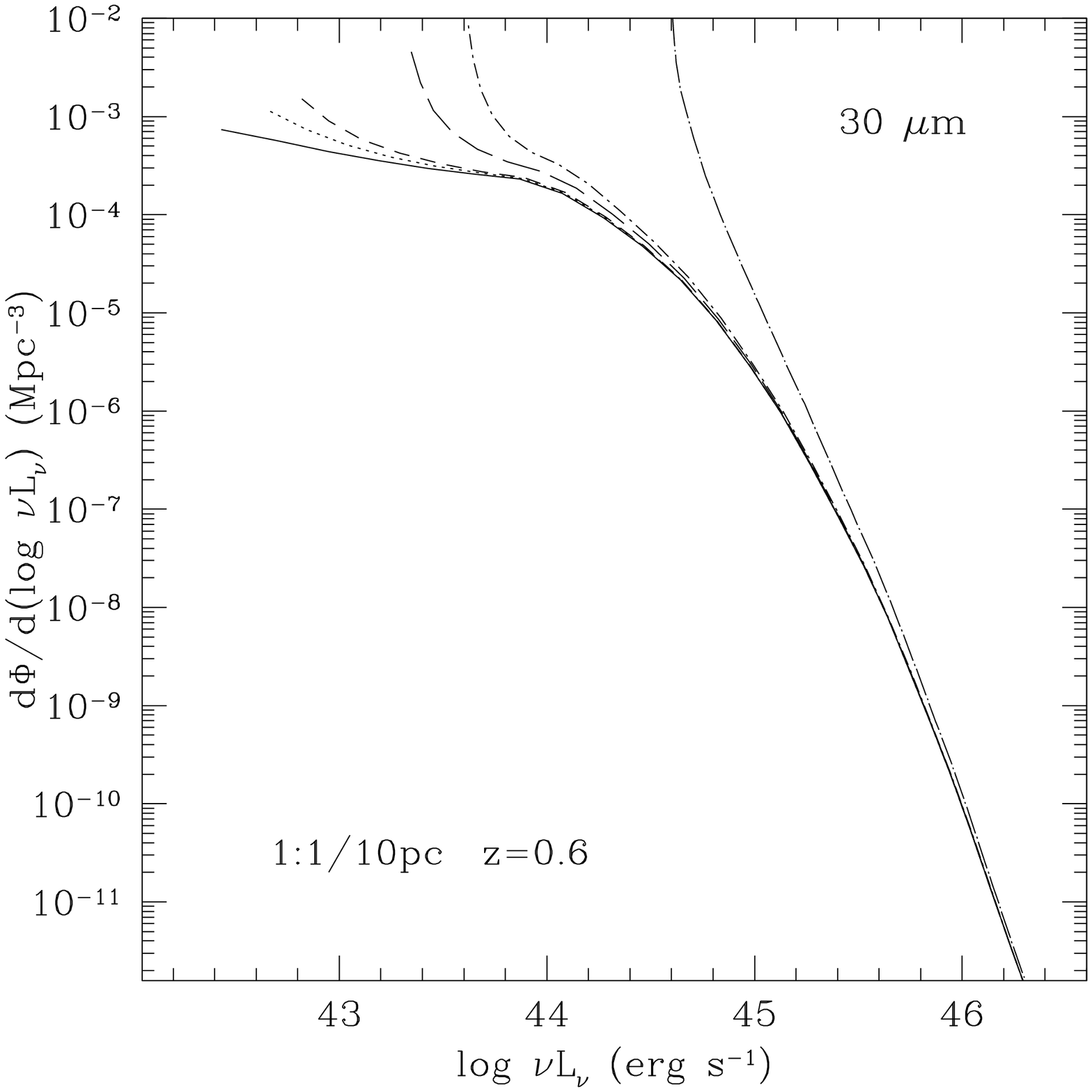}
\caption{The predicted rest-frame 30\micron\ luminosity functions from
  the 1:1 grids at $z=0.6$. The solid line has no contribution from
  star-formation and is identical to the equivalent curve in
  Fig.~\ref{fig:30umlf}. The other line styles show the prediction
  when a constant SFR is included using the dust emission spectrum of
  eqn.~\ref{eq:sf}: dotted-line=0.5~M$_{\odot}$~yr$^{-1}$,
  short-dashed line=1~M$_{\odot}$~yr$^{-1}$, long-dashed
  line=5~M$_{\odot}$~yr$^{-1}$, short dot-dashed
  line=10~M$_{\odot}$~yr$^{-1}$, long dot-dashed
  line=100~M$_{\odot}$~yr$^{-1}$. The left(right)-hand panel contains the
  results when the inner radius of the attenuating material is at
  1(10)~pc from the AGN.} 
\label{fig:30umlf-sf}
\end{figure}

\begin{figure}
\epsscale{1.0}
\plottwo{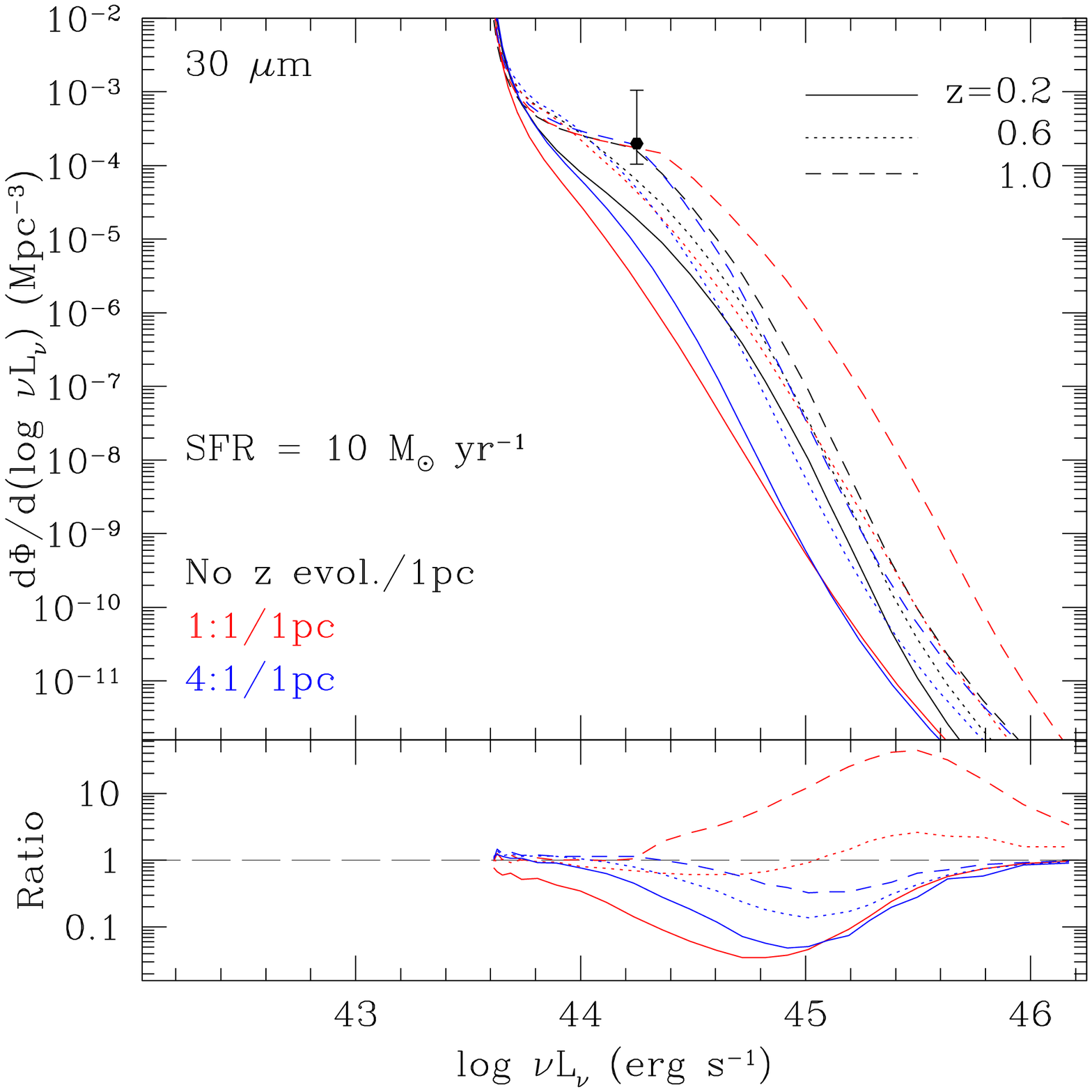}{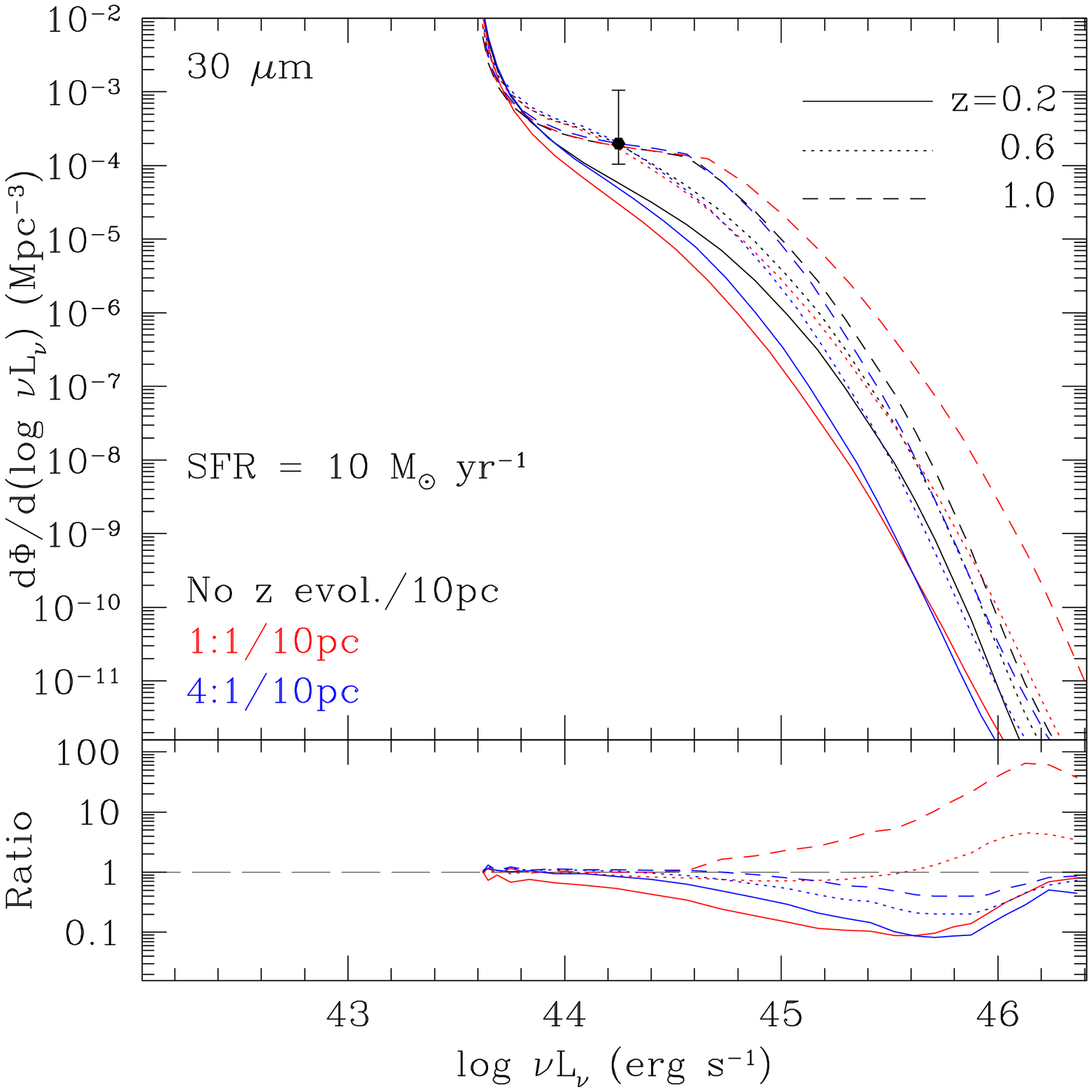}
\caption{As in Fig.~\ref{fig:30umlf}, but now including dust emission from a
  constant SFR of 10~M$_{\odot}$~yr$^{-1}$.}
\label{fig:30umlf-sf2}
\end{figure}

Assuming the 10~pc models are a more accurate guide, then the
right-hand panel of Fig.~\ref{fig:30umlf} indicates that largest
differences among evolutions of $R$ occurs at luminosities with
$\log \nu L_{\nu} > 45$. This is past the knee of the LF, and so will
require wide surveys to pick up the large numbers of luminous AGN. The
observations will also require significant depth in order to trace the
LF out to $z \sim 1$.

Another issue in observationally testing the LF predictions is
accessible wavebands. To really understand the evolution of the
obscuring region, one would like to trace the LF from low redshift
(say, $z=0.2$) to $z \sim 1$. The LF predictions are in the
rest-frame, which means that for 30\micron\ observations at $z \sim
0.6$, the required flux is being radiated at 48\micron. One would then
be forced to perform multi-wavelength observations and SED fitting for
each object in order to predict the flux at the required
wavelength. This SED fitting could be relatively simple as the largest
uncertainties for determining the LF will be in the space density
statistics.

Given these difficulties, it is natural to ask if the mid-IR LF is the
easiest method to determine the evolution of $R$. A perhaps more
straightforward strategy would be to measure the ratio of narrow-line
to broad-line AGN in the optical as functions of $z$ and
$L_{\mathrm{[O\ III]}}$, the luminosity in the [O III] $\lambda$5007
line. Alternatively, we could measure the ratio of X-ray AGN with
absorbing column densities $\log N_{\mathrm{H}} < 22$ to those with
$\log N_{\mathrm{H}} > 22$ as a function of $z$ and $L_{X}$, although an
accurate value of \nh\ requires an X-ray spectrum with sufficient counts. Since it is currently difficult to obtain enough
counts for spectral fitting from faint sources, hardness ratios
(which assume a spectral slope) are often used as an
alternative. Thus, both of these other methods to track the evolution
of $R$ require serious observational investments, but, more
problematically, they also are both subject to severe selection
effects. As is becoming increasingly apparent, optical selection of
AGN (i.e., selection on the basis of high-ionization lines) has
significant limitations \citep[e.g.,][]{netz06} and many (mostly type 2 AGN) are missed due to
obscuration or dilution \citep{mfc02,eck06,rig06}. Likewise, X-ray
selection is limited because the observable passband for the most
sensitive telescopes ends at $\sim 10$~\kev, and the observations are
thus insensitive to objects with high column densities, especially
those with \nh $> 10^{24}$~cm$^{-2}$, the Compton-thick AGN. The high
energy detectors on \textit{Swift} and \textit{Integral} have been
able to conduct surveys, but are limited to the nearby Universe due to
low sensitivity \citep{mark05,bas06}. Future observations with the
Hard X-ray Detector on board \textit{Suzaku} may help, but it is a
non-imaging instrument. The cancellation of the hard X-ray imaging
mission \textit{NuStar} makes finding significant numbers of the
Compton-thick AGN by their hard X-ray emission practically impossible
until the launch of \textit{Constellation-X}.

The advantage of the mid-IR method described here is that it is less
affected by the above selection effects. The predicted LFs shown in
Fig.~\ref{fig:30umlf} are for both type 2 and type 1
AGN, therefore detailed optical emission line spectra are not
required. Compton-thick objects should also be found in the infrared,
as the absorbed X-ray radiation will be re-emitted by the dust. The
difficulty, of course, is determining which IR sources are
predominantly powered by AGN as opposed to star-formation, although
methods that examine the spectral shape of the near- and mid-IR have
had some success \citep{ah06,brand06}, and AGN are well known to have
very weak PAH emission \citep{weed05}. However, X-ray selection is
still the best way to identify AGN, as Compton-thin objects can be
observed to high redshifts. Deep infrared observations are then
required to find the Compton-thick sources. Comparisons between
multiband observations and simulations similar to those performed here
will be useful in determining the AGN contribution to these potential
Compton-thick sources. This can be done by adding in contributions
from the host galaxy to the photoionization models, and computing a
large grid of IR properties over a range of AGN luminosities and
redshifts for comparison to both near- and mid-IR data.

Another interesting effect seen in the IR and not in the other
wavelengths is the dependence on the geometry of the obscuring
medium. Figures~\ref{fig:8umlf} and~\ref{fig:30umlf}
have shown different shapes for the LF depending on the inner radius
of the absorbing gas and dust. For the one data point at 30\micron, it
was difficult to determine the best evolution of $R$ without knowing
the geometry. Fortunately, there are significant differences in the
predicted LFs and number counts for the different inner
radii. Therefore, it is possible to break the degeneracy between $r$
and $R$, as was done at rest-frame 8\micron\ (Fig.~\ref{fig:8umlf}).

Finally, the mid-IR LFs presented above are relatively insensitive to
the actual \nh\ distribution assumed. We considered both a `simple'
\nh\ distribution, where all column densities had equal weight within
the type 1 or 2 designations (eqn.~\ref{eq:simple}), and the
\citet{rms99} \nh\ distribution in which half of the type 2 AGN are
Compton thick and 75\% had $\log N_{\mathrm{H}} > 23$
(eqn.~\ref{eq:risaliti}). These weights were used in constructing both
the `unified' and \nh-averaged SEDs. Despite this, the two
distributions made only negligible differences to the predicted number
counts and LFs. This is because the additional weight afforded to
Compton thick AGN in the \citet{rms99} \nh\ distribution has only a
small effect in the mid-IR (Fig.~\ref{fig:sedexample}). Therefore,
although mid-IR LFs will not be able to quantify the relative number
of Compton thick AGN, they are still very sensitive to the overall
evolution of the AGN type 2/type 1 ratio.

\subsection{The Role of Star Formation}
\label{sub:stars}
Star formation is a key process in this investigation in two distinct
ways. First, it acts as a contaminant of the mid-IR emission from the
obscured AGN, masking the evolutionary effects we are trying to
investigate. Using a straightforward estimate of the effects of star
formation on the mid-IR, we found in Sect.~\ref{sect:sf} that for a
constant SFR $\leq 10$~M$_{\odot}$~yr$^{-1}$ the impact at 30\micron\
was confined to $\log \nu L_{\nu} < 44.5$, which translates to X-ray
luminosities of $\log L_{X} \lesssim 43.5$ for the 1:1/10pc models or
$\log L_{X} \lesssim 43.7$ for the nozevol/10pc and 4:1/10pc
grids. The affected luminosities are low enough that the 
 rest-frame 30\micron\ LFs may still be used to discriminate possible
 different evolutions of the AGN type 2/type 1 ratio
 (Fig.~\ref{fig:30umlf-sf}). The method may prove more problematic
 with galaxies with average SFRs $> 10$~M$_{\odot}$~yr$^{-1}$.

The sources that produce the majority of the CXRB have been
identified to be predominantly obscured AGN with $43 \leq \log L_{X}
\leq 44$ \citep{bar05}. These objects are typically found at $z
\lesssim 1$, which may signify a relationship between the obscuring
material around an AGN and the star formation in the host galaxy
\citep{fab98,fra99,bem06}. It is this possibility that makes measuring
the evolution of $R$ an important goal in the study of galaxy
evolution. However, there may be a potential difficulty in employing
the mid-IR LFs calculated here because, as already noted, the
important luminosity range to distinguish the various evolutions is
$\log L_{X} \gtrsim 44$, larger than that of the objects that dominate the
X-ray background. The reason for this is that the three evolutions
considered have very different predictions for the ratio of type
2/type 1 quasars (Fig.~\ref{fig:contours}), but quasars probably have
a very different formation and evolution history than the lower
luminosity Seyferts that comprise the bulk of the CXRB. Quasars are
more commonly found at $z \gtrsim 1$, are likely formed by the mergers
of massive galaxies, and simulations show their host galaxies have
little ongoing star formation within the galaxy at late times
\citep[e.g.,][]{hopk05}. The evolutions of $R$ described by
eqns.~\ref{eq:4to1} and~\ref{eq:1to1} assume that the AGN covering
factor increases with $z$ independent of the AGN luminosity. If the
attenuating gas and dust is connected to the host galaxy SFR then this
may not be correct at large $L_{X}$, and indeed very tentative
evidence for a slower increase of $R$ with $z$ at larger $L_{X}$ was
found by \citet{bem06}. If this is the case, then evidence for this
type of luminosity-dependent redshift evolution will appear in the 30\micron\ LFs, and can be measured using the techniques presented
here. Furthermore, the effects of our luminosity-independent
assumption are most important at high $z$, but Fig.~\ref{fig:30umlf} show that significant variations in the
rest-frame 30\micron\ LF due to the different evolutions of $R$
still occur at low redshifts. Thus, a determination of the LF at
redshifts $z \sim 0$--$0.2$ will result in an important constraint on
the local value of $R$. Then, tracking the 30\micron\ LF with
redshift will yield a measurement of how the AGN type 2/type 1 ratio
changes with $z$ and $L_{X}$, a key parameter in understanding the
formation and evolution of these galaxies.

\subsection{Testing the Unified Model}
\label{sub:constantcovering}
To this point, all the calculations presented in this paper
assumed that the unified model of AGN was correct. That is, the
fraction of type 2 AGN at a given $L_X$ and $z$ is equal to the
covering factor of the absorbing material around the AGN. These
models can also be used to test whether the unification scenario is valid
at all luminosities. 

As an example, we searched for \textit{IRAS} objects satisfying
$|b|>35$ degrees, $f_{\nu}$(60\micron)$>0.2$~Jy and with redshifts
between 0.8 and 1.2. This results in 13 objects in which we identified
8 quasars using the literature. Two of them ([HB89]0235+164 and
[HB89]1308+326) are blazars and are omitted from further
consideration. An additional 3 (IRAS F09121+2430, PG1248+401 and
B31340+407) were removed because they have weak detections at
60\micron\ and the association with the quasars therefore is not
firmly established. The remaining three quasars
(2MASSiJ0119560-201022, 2MASSiJ1543519+162422 and PG1206+459) have
strong detections by \textit{IRAS} and the optical data show them to
have normal QSO SEDs.  Assuming $z \sim 1$, the rest-frame 30\micron\
luminosities of these quasars are $\nu L_{\nu} \sim
46.8$. Fig.~\ref{fig:30umlf} shows that the 1:1/10pc evolutionary
model, as a result of its $\sim$3:1 type 2/type 1 ratio
(Fig.~\ref{fig:contours}), can reach a 30\micron\ luminosity of $\nu
L_{\nu} = 46.5$ at $z=1$, but this occurs only when the input X-ray
luminosity is $\log L_{X} = 48$. Yet, the 2--10~\kev\ X-ray luminosity
of PG1206+459 is $\sim 10^{45}$~erg~s$^{-1}$ \citep{pic05}. Indeed,
using the three \textit{IRAS} sources to estimate the 30\micron\ LF at
$\nu L_{\nu} = 46.8$ results in a lower-limit of
$>10^{-10}$~Mpc$^{-3}$, about 2 orders of magnitude larger than the
predicted LF from Fig.~\ref{fig:30umlf}. The best way of reducing this
discrepancy would be if the models could produce these high IR
luminosities at lower AGN input luminosities.

\begin{figure}
\epsscale{1.0}
\plottwo{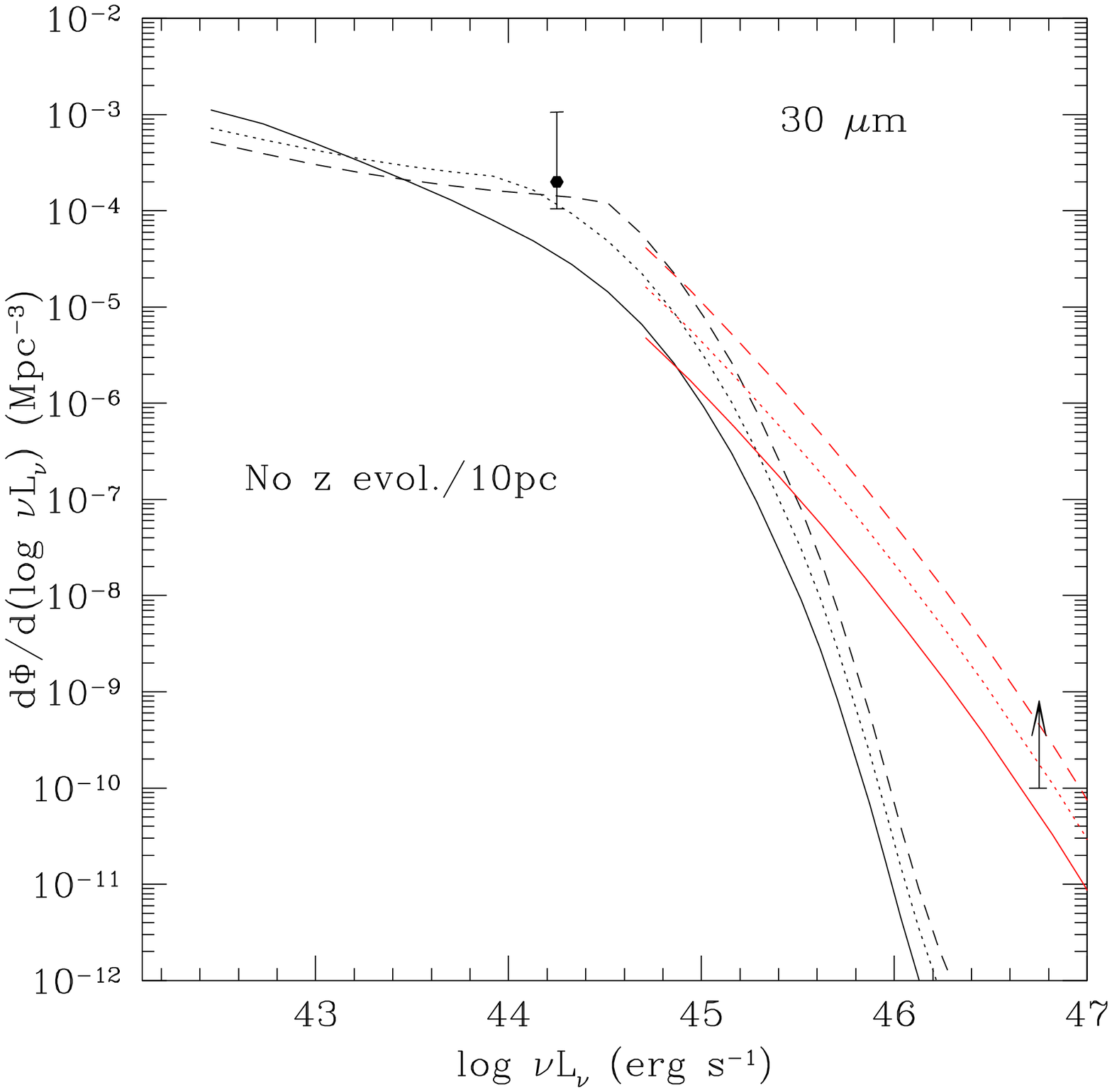}{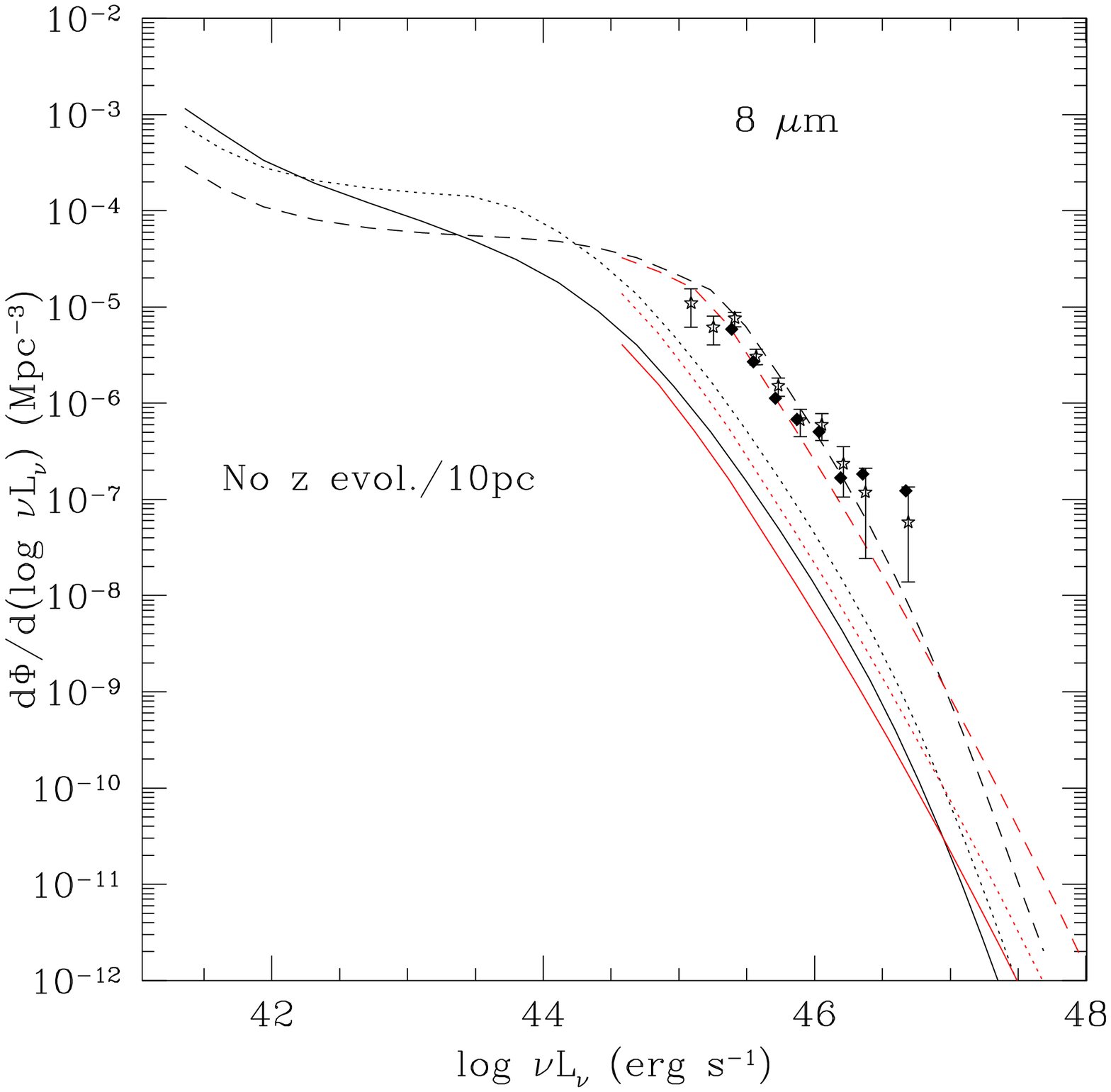}
\caption{(Left) Plot of the rest-frame 30\micron\ LF at $z=0.2$, $0.6$
and $1.0$ (line styles and low-luminosity data point as in
Fig.~\ref{fig:30umlf}). The LFs
were computed using the nozevol/10pc model grid. The arrow denotes the
lower-limit to the high-luminosity rest-frame 30\micron\ LF at $z=1$
derived from \textit{IRAS} sources (see text). The black lines are
identical to the corresponding ones from Fig.~\ref{fig:30umlf} and plot the
predictions using the assumed compact geometry described in
Sect.~\ref{sub:assume}. The red lines show the high-$\nu L_{\nu}$ results when the 
density of the absorbing cloud is lowered to 100~cm$^{-3}$. The
absorber extends to much greater distances than the previous case, and
therefore enhances the emission in the mid-IR. (Right) As in the the
left-hand plot, but for the rest-frame 8\micron\ LF. The line styles
and points are the same as in Fig.~\ref{fig:8umlf}.}
\label{fig:iras}
\end{figure}

It seems that the assumed geometric configuration of a relatively
compact absorbing medium situated at 1~pc or 10~pc from the AGN cannot
produce the necessary 30\micron\ emission at high $L_X$ to account for
the \textit{IRAS} limit. Recall that because of the assumed density of
10$^4$~cm$^{-3}$, the pathlength through the absorber is only $15$~pc
for $\log N_{\mathrm{H}} = 23$ and $\log L_X = 48$ (taking into
account the ionization effects), therefore at high AGN luminosities
there is only a small column of dust cool enough to emit at
30\micron. Material at greater distances is needed to increase the
30\micron\ emission. A simple way to test this idea is to calculate a
new grid of \cloudy\ models with the inner radius remaining at 10~pc
but the density now lowered to 10$^2$~cm$^{-3}$. This greatly
increases the pathlength of the absorbing material (it is now over
1~kpc in the example above). Figure~\ref{fig:iras} plots the
rest-frame 30\micron\ LFs computed using the models with a lower
density absorber as well as the previous derived ones that assume the
more compact geometry. Only results for $\log \nu L_{\nu} \gtrsim
44.7$ are plotted for the low density absorber. The greater pathlength
afforded by the lower density cloud now produces enough 30\micron\
emission to account for the \textit{IRAS} lower-limit. The right-hand
panel of Fig.~\ref{fig:iras} shows the same results but for the
rest-frame 8\micron\ LF. At this wavelength, the differences in the LF
between the two assumed densities are small.

The implication of this simple test is that the high luminosity
quasars require absorption and remission of nuclear radiation over a
large range of radii, while the current data available for the lower
luminosity quasars and Seyfert galaxies can be adequately described by
a more compact absorber. The type 2/type 1 ratio at the \textit{IRAS}
luminosities for the three evolutions considered in this paper varies
from $\sim$1:1--4:1 (Fig.~\ref{fig:contours}), a range that is
consistent with the latest observational constraints
\citep{mart05,mart06}. Under the unification model, this translates
into covering fractions of 0.5--0.8. Producing high covering fractions
is significantly simpler if the obscuring material is close to the
AGN, and not spread out over a large range of radii, as implied by the
above results. Clearly, the change in mid-IR LF with luminosity (in
addition to the possible redshift evolution concentrated on in
previous sections) contains fundamental clues on the nature and
geometry of the obscuring medium and therefore the host galaxy. Future
observations are needed for this technique to fulfill this promise.

We conclude that the unification model based on a relatively compact
absorber is consistent with the available data for moderate AGN
luminosities (Seyfert galaxies) but fails when extending this to the
high luminosity quasars. This adds to the evidence that the evolution
and fueling of quasars are very different from those that drive the
lower luminosity Seyfert galaxies.

\section{Summary}
\label{sect:summ}
The principle conclusion of this work is that a measurement of the
redshift evolution of the mid-IR AGN LF will determine to what extent
the AGN type 2/type 1 ratio, $R$, is evolving with $z$. The specific
predictions shown above were of most use for the existing \spitzer\
MIPS bands, but are equally valid for future observations with
\textit{JWST}. Predictions were shown for three different evolutions
of $R$, but the current data were unable to distinguish among them.

The calculations performed here were relatively novel, as they
employed the photoionization code \cloudy\ to directly connect the
X-ray properties of AGN to the IR. This technique, although not
appropriate for fitting SEDs of individual objects, yielded
\nh-averaged SEDs that well described the average properties of AGN
found in the deep surveys. In addition, varying the parameters of the
models allows constraints to be placed on the physical conditions for
an average AGN absorber. For example, comparing results assuming two
different distances for the average inner radius of the attenuating gas and
dust (1~pc and 10~pc), yielded compelling evidence that the rest frame
mid-IR properties of a large ensemble of AGN
were best described when the inner radius of the obscuring material
was $\sim 10$~pc from the AGN. Future work
will cover a greater parameter space and split the data into both
redshift and luminosity bins. In this way, properties of average AGN
can be derived and followed with minimal interference from the
intrinsic object-to-object variability.

Finally, it is worth emphasizing the importance of determining the
evolution of the AGN type 2/type 1 ratio, as it may be directly
related to processes within the host galaxy. Between $z \sim 0$ and
$1$, most AGN are relatively low-luminosity Seyfert galaxies and
therefore such studies probe a different mode of galaxy formation and
evolution than the higher-luminosity quasars at high
redshifts. Evidence for such a difference was uncovered by estimating
the rest-frame 30\micron\ LF at high quasar luminosities. This LF was
more than two orders of magnitude above the predictions, which
assumed that a compact unification model was valid at all
luminosities. To account for this observational limit absorbing
material at much larger radii was invoked, suggesting a different mode
of obscuration. Seyfert galaxies, in contrast, can still be
successfully described by the compact unified model.

\acknowledgments

This research has made use of the NASA/IPAC Extragalactic Database
(NED) which is operated by the Jet Propulsion Laboratory, California
Institute of Technology, under contract with the National Aeronautics
and Space Administration.  DRB is supported by the University of
Arizona Theoretical Astrophysics Program Prize Postdoctoral
Fellowship. Partial support for this work was provided by NASA through
a contract 1255094 issued by JPL/Caltech. We thank G.J.\ Ferland for
providing the community with \cloudy, D.\ Psaltis for allowing us to
use the Sierra cluster for the computations, and J.\ Everett for
comments on a draft of the paper.

\clearpage

\clearpage

\clearpage

\clearpage

\clearpage

\clearpage

\clearpage

\clearpage

\clearpage

\clearpage

\clearpage

\clearpage

\clearpage

\end{document}